\documentclass[
    12pt,
    a4paper
]{article}

\usepackage[T1]{fontenc} % T1 gives a real Å glyph; without it \AA is composed and the ring sits off-centre
\usepackage{newtxtext}
\usepackage{newtxmath}

\usepackage{amsmath}

\usepackage{amssymb}

\usepackage{graphicx}
\usepackage{float}

\usepackage{booktabs}
\usepackage{array}

\usepackage[hidelinks]{hyperref}
\usepackage{setspace}

\usepackage{titlesec}

\titleformat{\section}
  {\normalfont\bfseries\large}
  {\thesection.}{0.5em}{}

\titleformat{\subsection}
  {\normalfont\bfseries\normalsize}
  {\thesubsection.}{0.5em}{}

\titlespacing*{\section}
  {0pt}      % Einzug
  {2.5ex}    % Abstand VOR der Section
  {2.5ex}    % Abstand NACH der Section

\titlespacing*{\subsection}
  {0pt}{2ex}{2ex}

\usepackage{graphicx}% Include figure files
\usepackage{subcaption}
\usepackage{pgfplots}
\pgfplotsset{compat=1.18}
\usepackage{pgfplotstable}
\usepackage{dcolumn}% Align table columns on decimal point
\usepackage{bm}% bold math
\usepackage[hidelinks]{hyperref}
\usepackage{color} %to color the text
\usepackage{siunitx}
\DeclareSIUnit\eangstrom{\text{e\AA}}
\usepackage{amsmath}
\usepackage{braket}
\usepackage{mathtools}
\usepackage[version=4]{mhchem} %for chemical formulas

\usepackage[normalem]{ulem}
\usepackage[shortlabels]{enumitem}
\usepackage{enumerate}
\usepackage{enumitem}
\usepackage{upgreek}
\usepackage{booktabs}
\usepackage{cleveref}
\crefname{figure}{Fig.}{Figs.}\Crefname{figure}{Fig.}{Figs.}
\crefname{subsection}{Sec.}{Secs.}\Crefname{subsection}{Sec.}{Secs.}
\crefname{section}{Sec.}{Secs.}\Crefname{section}{Sec.}{Secs.}
\usepackage{ragged2e}

\usepackage[
    a4paper,
    margin=1.8cm,
    columnsep=0.6cm
]{geometry}

\usepackage[
    backend=biber,
    style=numeric-comp, % [4]-[6] instead of [4][5][6]; use \cite{a,b,c}
    sorting=none,
    isbn=false
]{biblatex}
\AtEveryBibitem{%
    \ifentrytype{article}{\clearfield{title}}{}%
    \ifentrytype{inproceedings}{\clearfield{title}}{}%
    \clearfield{doi}%
}

\begin{document}

\textbf{Photonic and Electronic Interaction of TMD-Hybridized ZnO Nanowire Lasers}
\\
\\
Moritz Willems*$^a$, Edwin Eobaldt*$^a$, Alexander Zaunick$^a$, Aurelia Eberhard$^a$, Maximilian Tomoscheit$^a$, Lukas Raam Jäger$^a$, Xiao Chen$^b$, Preeti Sharma$^b$, Nicolas Tancogne-Dejean$^b$, Antony George$^c$, Andrey Turchanin$^{c,d}$, Silvana Botti$^b$, Giancarlo Soavi$^{a,d}$ and Carsten Ronning$^{a,d}$
\newline

\noindent
$^a$Institute of Solid State Physics, Friedrich Schiller University Jena, 07743 Jena, Germany
\newline
$^b$RC Future Energy Materials and Systems $\&$ ICAMS, Faculty of Physics, Ruhr University Bochum, 44801 Bochum, Germany
\newline
$^c$Institute of Physical Chemistry, Friedrich Schiller University Jena, 07743 Jena, Germany
\newline
$^d$Abbe Center of Photonics, Friedrich Schiller University Jena, 07745 Jena, Germany
\newline
\newline
*These authors contributed equally to this study.
Corresponding authors: moritz.willems@uni-jena.de, carsten.ronning@uni-jena.de

Keywords: semiconductor nanowires, heterostructures, lasing, FDE simulations, DFT calculations

\newpage

\begin{abstract}
Semiconductor nanowire lasers are promising on-chip coherent light sources for photonic integrated circuits, but integration inevitably places them in contact with other materials, thereby changing their properties upon hybridization. 
In this study, we examine the photonic and electronic coupling between single ZnO nanowire lasers and TMDs. 
Using deterministic nanomanipulation, we first hybridized the same nanowire first with a WS$_2$ monolayer and then with a MoS$_2$ monolayer. 
The respective lasing threshold rises, the emission is strongly quenched at high pump power densities, and the mode envelope blueshifts. 
Upon hybridization with bilayer, trilayer, and bulk MoS$_2$, lasing is fully suppressed. 
Finite-Difference Eigenmode simulations show that the TMD absorbs part of the evanescent field. 
Because this absorption is wavelength dependent, it also removes modes from the low-energy flank of the envelope, which accounts for small difference between the two monolayers, while the blueshift itself follows the increase of the carrier density at threshold.
Density functional theory calculations that include the interface dipole predict a type-I alignment at the ideal ZnO(10$\bar{1}$0)/TMD interface, which turns into type~II for a moderately reduced dipole, as expected for hydroxylated facets in air. In both cases, holes are transferred from the nanowire into the TMD and removed from the gain medium. Evanescent absorption and hole transfer thus jointly suppress lasing, and both must be accounted for when integrating nanowire lasers with 2D materials.
\end{abstract}

\newpage

\section{Introduction}

Consistent with Moore's law, integrated circuits have enabled exponential advances in computing over the last few decades.\cite{Moore.2006} As transistors have become smaller and smaller, down to the sub-10-nm scale, quantum tunneling effects occur, creating leakage currents.\cite{Tsang.2025} Furthermore, high power densities cause increasingly strong thermally induced stresses which can negatively affect the reliability, lifetime, and performance of semiconductor devices.\cite{Ibrahim.2024} 
Photonic integrated circuits (PICs) have emerged as a promising technology to complement advanced electronic integrated circuits (EICs) by using photons as information carriers, mitigating some fundamental limitations of the EICs.\cite{Bogaerts.2020, Li.2019, HuaTan.2023} While electronics remain essential for high-density computation, photonic systems can complement electronic circuits by routing high-speed data streams with excellent energy efficiency and low signal degradation.\cite{Kaur.2021} Beyond information technology, the integration of photonic components also enables compact lab-on-a-chip platforms for applications such as label-free detection of biological targets, cancer diagnostics, and on-chip disease detection.\cite{RamirezPriego.2026, Zhang.2017}

A key requirement for such integrated photonic systems is the availability of efficient and controllable on-chip coherent light sources.
Semiconductor nanowires can serve as nanolasers and, when excited, provide a coherent beam of photons.  They have the intrinsic properties necessary for a laser, except an excitation source: they act as a gain medium, a waveguide, and a Fabry–Pérot resonator.\cite{Zimmler.2010, Ma.2013} ZnO nanowires are of particular interest, which provide room-temperature lasing due to their strong exciton binding energy and a stable electron-hole plasma.\cite{Vanmaekelbergh.2011, Huang.2001, Johnson.2001}

When integrated into a PIC, a nanowire comes into contact with other materials and is exposed to hybridization effects. Understanding the photonic and electronic coupling mechanisms behind this is fundamental to their integration, because these interactions can substantially alter light-matter coupling. Previous research has shown how the hybridization with metals increases the speed of the lasing emission.\cite{Eaton.2016, Roder.2017}
Research regarding the interaction of lasing nanowires with TMDs is scarce, even though TMDs, such as MoS$_2$ and WS$_2$, exhibit strong excitonic characteristics and a direct band gap in the monolayer limit, making them particularly promising for modifying the optical properties of nanowire lasers.\cite{tomoscheit2026twist} 
In our previous study, we investigated the interaction of nanowires with monolayer MoS$_2$ and observed a partial suppression of the lasing emission.\cite{Eobaldt.2022} We assumed a type-II band alignment between ZnO and MoS$_2$ \cite{Liu.2016} as well as an associated charge transfer process. Thus, we attributed the change in the lasing characteristics to band filling and the Moss-Burstein effect. 
However, we did not consider further loss mechanisms for the lasing emission, such as optical absorption by the 2D material being in the evanescent field of the lasing nanowire.

In this study, we therefore additionally investigated the lasing characteristics of ZnO nanowires hybridized with WS$_2$ monolayers that possess a different valence-band offset relative to MoS$_2$. We examined the emission properties of the hybrid structures as a function of the thickness of the 2D material, and focused on the photonic coupling in the multidimensional hybrids. Based on these experimental results, we revisited the electronic coupling using density functional theory (DFT) calculations to develop a consistent picture of the photonic and electronic coupling mechanisms between ZnO nanowires and 2D materials.

\section{Results and Discussion}
\subsection{Lasing of TMD monolayer-hybridized ZnO nanowires}
\label{sec:lasing}

\begin{figure}
    \centering
    \includegraphics[width=\linewidth]{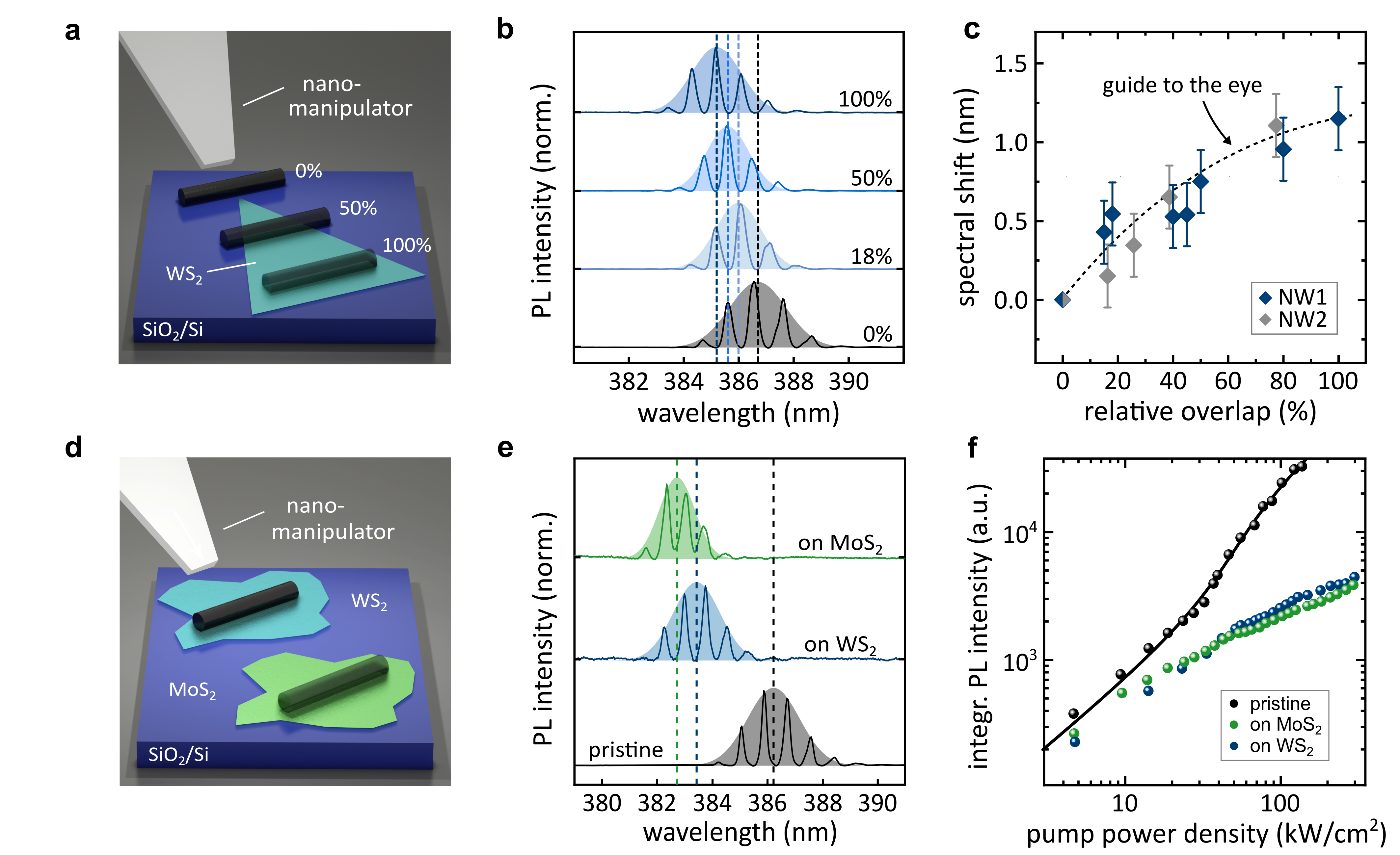}
    \caption{\justifying \textbf{Hybridized nanowire lasing on WS$_2$ and MoS$_2$ monolayers.} (a) Schematic view of the deterministic hybridization of a ZnO nanowire with a WS$_2$ monolayer, enabling controlled overlap from 0\% to 100\%. (b) Normalized µPL spectra at high excitation ($\sim$\,65\,kW/cm$^2$), showing a progressive blueshift and mode evolution with increasing overlap. (c) Lasing envelope peak shift as a function of overlap for two individual nanowires on WS$_2$. (d) Schematic view of the same nanowire sequentially hybridized with WS$_2$ and MoS$_2$ monolayers. (e) Normalized µPL spectra in the lasing regime for pristine, WS$_2$-, and MoS$_2$-hybridized configurations. (f) Integrated PL intensity versus pump power density for pristine and TMD-hybridized nanowires; Casperson fit to the pristine nanowire (solid line).}
    \label{Results1}
\end{figure}

Both ZnO nanowires and TMD monolayers were placed next to each other on a SiO$_2$/Si substrate using the polydimethylsiloxane (PDMS) transfer technique.\;A nanomanipulator was then used to mechanically move the nanowires on top of the TMD monolayers, thereby yielding a range of overlap ratios, as schematically illustrated in \cref{Results1}a. 
This experimental scheme allows the collection of the lasing spectra  of the same individual ZnO nanowire, excluding any variation of the optical data observed for varying geometries of nanowires, which was a limiting issue in our previous study.\cite{Eobaldt.2022} 

First, we hybridized one ZnO nanowire with a WS$_2$ monolayer and pumped it into the lasing regime at $\approx$\,65\,kW/cm$^2$. The lasing spectra as a function of the relative overlap are shown in \cref{Results1}b. 
With increasing overlap, the mode envelope shifts clearly to shorter wavelengths. The shift for this nanowire, plotted against the overlap in \cref{Results1}c, reaches about 1.2\,nm at full overlap. A second nanowire behaves in the same way, as shown in \cref{Results1}c, and the shift is similar to that reported for MoS$_2$ hybridization.\cite{Eobaldt.2022} 

In another experiment, we mechanically moved an individual nanowire off the WS$_2$ monolayer after collecting the optical data. Then, we transferred the nanowire fully onto a MoS$_2$ monolayer (see the schematic \cref{Results1}d). 
The two TMDs have similar optical band gaps (2.01\,eV for WS$_2$ and 1.88\,eV for MoS$_2$) but different valence band offsets to ZnO.\cite{Dai.2025} 
This may cause faster charge-transfer dynamics and a stronger influence on the lasing characteristics. 
It is indeed evident that both TMDs induce a slightly different blueshift of the lasing spectra of the ZnO nanowires, as shown in \cref{Results1}e.
Upon hybridization with WS$_2$, a blueshift of $\sim$\,2.8\,nm is observed. In the case of MoS$_2$, the emission maximum shifts by $\sim$\,3.2\,nm, with an uncertainty of 0.2\,nm each. 
This observation fits with the expectation of a faster charge transfer dynamics, causing higher losses for the lasing ZnO nanowire and the need for stronger band filling to maintain the lasing conditions of an electron-hole plasma. Furthermore, a closer look at the lasing modes reveals that the high-energy cutoff of the mode envelope remains unchanged across both hybrid configurations, whereas the MoS$_2$-hybridized nanowire supports fewer low-energy modes. This reduced support in the MoS$_2$ configuration results in a narrowing of the mode envelope, which in turn leads to the observed shift in the central emission wavelength.

Furthermore, we plot in \cref{Results1}f the integrated photoluminescence (PL) intensity of the ZnO nanowire as a function of the pump intensity for the pristine nanowire, as well as after hybridization with the two different TMD materials. The lasing threshold can be determined from such light-in-light-out (LL) curves by fitting them to a multimodal lasing model developed by Casperson.\cite{Casperson.1975} For the pristine ZnO nanowire, a characteristic S-shaped curve is evident and the fit yields a lasing threshold of (43\,$\pm$\,3)\,kW/cm$^2$.\cite{Eobaldt.2026}
Hybridization significantly reduces the emission intensity, especially at high pump powers, resulting in a strong deviation of the S-shaped curvature. Notably, the emission curves of the MoS$_2$ and WS$_2$ hybridized nanowires do not differ significantly and the threshold for both hybrid structures were determined to be approximately 95\,kW/cm$^2$. The two threshold values agree within the experimental uncertainty despite the different band offsets. Therefore, in addition to charge-transfer losses, another loss mechanism must be present. We identified this mechanism as optical absorption by the 2D material, which overlaps with the evanescent field of the nanowire.

\subsection{Layer-dependent optical response of MoS$_2$-nanowire hybrid systems}
\label{sec:layers}

Different ZnO nanowires were transferred onto MoS$_2$ flakes of varying thickness, from monolayer to bulk, using the PDMS transfer technique. For comparability, all nanowires were taken from the same growth batch and had similar diameters (200-280\,nm) and lengths (10-15\,µm). The number of layers was determined via Raman spectroscopy, by measuring the frequency difference between the $E^1_{2g}$ and $A_{1g}$ modes, 
as well as photoluminescence measurements showing an increasing redshift of the excitonic transitions and a decrease in emission intensity for higher layer numbers.\cite{Bing.2018} (Supporting Information, Fig.~S1). 

\begin{figure}
    \centering
    \includegraphics[width=\linewidth]{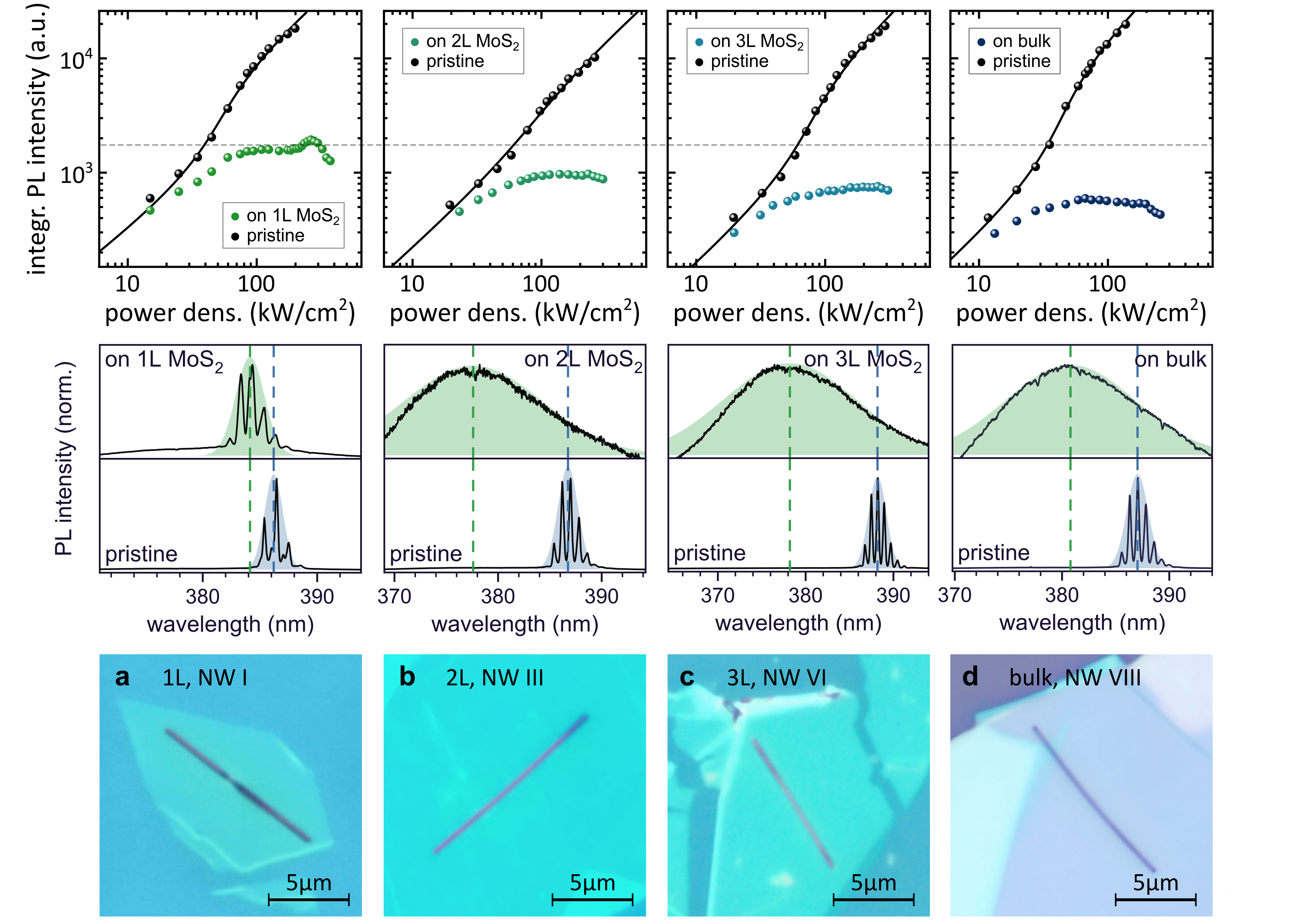}
    \caption{\justifying \textbf{Layer-dependent optical response and threshold breakdown in hybrid $\text{MoS}_2$-nanowire systems.}
    (top row) Log-log PL intensity vs.\;pump power density is shown for the pristine nanowires and after hybridization with flakes of varying thickness (1L, 2L, 3L, and bulk; all nanowires from the same growth substrate).\;(middle row) Photoluminescence spectra with excitation powers high enough for the pristine nanowire to lase are shown for the different hybridizations.\;(bottom row)Optical microscope images of the individual ZnO nanowires on $\text{MoS}_2$.} 
    \label{Results2}
\end{figure}

In \cref{Results2}a-d, we plot the optical response of hybridized ZnO nanowires with increasing TMD thickness from monolayer to bilayer, trilayer and bulk MoS$_2$. The top graphs display the LL curves for the different hybrid systems, the respective middle panels show the PL spectra for pump powers deep in the lasing regime of the pristine nanowire, and the bottom panels show the corresponding optical microscope images.
The nanowire hybridized with monolayer MoS$_2$ in \cref{Results2}a shows an LL curve similar to the one shown in \cref{Results1}f, with slight deviations caused by differences in the nanowire's morphology.
The emission is only weakly quenched at low pump power densities, in the spontaneous emission regime, while above 50\,kW/cm$^2$ the S-shaped curve is again strongly flattened. Despite the quenching, the underlying photoluminescence spectrum still shows that the hybridized nanowire exhibits lasing emission, albeit with the envelope shift, as discussed previously.

A bilayer quenches the emission further and leads to saturation at even lower power densities (\cref{Results2}b). Moreover, the photoluminescence spectra of the bilayer hybrid contain no lasing modes and show only the broad near-band-edge emission band, i.e., stimulated emission is fully suppressed. Unlike for the monolayer, no lasing appeared even when pumping up to the degradation threshold of the nanowire.\cite{Zapf.2017}
These suppression characteristics can also be seen in \cref{Results2}c, d for trilayer and bulk MoS$_2$. The LL curves are suppressed even more strongly and saturate at lower pump power densities, and the PL spectra show no lasing at any excitation power.

While the band gap of MoS$_2$ shrinks when adding layers, its valence-band maximum shifts towards the vacuum level, so that hole transfer from the nanowire into the TMD remains energetically favorable for all number of layers. The stronger suppression for thicker flakes therefore points first to absorption: the thicker the TMD, the larger the share of the evanescent field of the guided modes that it absorbs.
This absorption matters most in the near UV, where the ZnO emission overlaps with strong optical resonances of the TMDs.
Besides the A and B excitons in the visible range, MoS$_2$ and WS$_2$ both show the so-called C and D resonances in the near UV.\cite{Song.2019,Ermolaev.2020,Jung.2018} They originate from parts of the Brillouin zone in which the conduction and valence bands exhibit almost parallel dispersion (band nesting).\cite{Carvalho.2013,Mennel.2020} This strongly raises the joint density of states and therefore the absorption at the emission wavelength of ZnO. At 385\,nm, monolayer absorption coefficients of about $9\times10^{5}$\,cm$^{-1}$ for WS$_2$ \cite{Jung.2018,Ermolaev.2020} and about 10$^{6}$\,cm$^{-1}$ for MoS$_2$ \cite{Song.2019,Ermolaev.2020,Jung.2018} have been reported. The monolayer thickness of only 0.7\,nm \cite{Mak.2010} limits the resulting absorption. Still, a part of the evanescent field of the lasing modes is absorbed and adds to the optical losses.

\subsection{Simulation of the photonic interaction}
\label{sec:photonic}

We quantified these absorption losses with Finite-Difference Eigenmode (FDE) simulations of a 250\,nm thick ZnO cylinder lying on top of a 0.7\,nm WS$_2$ monolayer on a SiO$_2$ substrate. The refractive indices and extinction coefficients of all materials were taken from the literature.\cite{Song.2019,Ermolaev.2020,Arosa.2020,Wille.2016} Nanowires of this thickness support several radial modes that compete for gain. 
The TE$_{01}$ mode has the highest gain and is mainly responsible for lasing,\cite{Zimmler.2010,Roder.2016} but its  electric field maximum lies close to the nanowire surface,\cite{Henneghien.2009} so it also suffers the highest propagation loss, which we find to be $\sim$\,1150\,cm$^{-1}$ with the WS$_2$ monolayer, as shown in \cref{Results3}b. The fundamental HE$_{11}$ modes, shown in \cref{Results3}a, in contrast, have their electric field maxima in the nanowire core and extend much less into the surroundings.\cite{Henneghien.2009} 
The substrate splits this mode into two polarizations, HE$_{11,x}$ and HE$_{11,y}$,\cite{Henneghien.2009} whose losses did not differ significantly in the simulations; we therefore report their average, which amounts to $\sim$\,390\,cm$^{-1}$ and is far below the TE$_{01}$ value. 
A pristine ZnO nanowire with a diameter of 200-300\,nm offers a maximum modal gain of 1600-2800\,cm$^{-1}$ at room temperature.\cite{Richters.2012} 
The absorption of the TMD in the evanescent field, especially of the TE$_{01}$ mode, consequently takes up a large part of the available gain.
Replacing WS$_2$ with MoS$_2$ in the otherwise identical structure gives modal losses of 1000\,cm$^{-1}$ for the TE$_{01}$ mode and 340\,cm$^{-1}$ for the HE$_{11}$ mode. The losses are thus slightly lower for MoS$_2$, but the difference between the two materials is small.

We repeated the simulations for multilayer MoS$_2$ flakes.\;At the emission wavelength, a bilayer gives 1850\,cm$^{-1}$ (TE$_{01}$) and 620\,cm$^{-1}$ (HE$_{11}$), nearly twice the monolayer values, and a trilayer gives 2500\,cm$^{-1}$ and 850\,cm$^{-1}$, respectively.
Absorption therefore contributes strongly to the suppression of stimulated emission in hybridized nanowires as the number of TMD layers grows. It cannot, however, explain why the emission curve saturates so markedly at high pump power density.\;A possible reason is that thicker TMD flakes allow faster carrier diffusion, which provides more states that can accept charge at the heterointerface. This mechanism could also account for the partial recovery of emission of the monolayer-hybridized nanowire in \cref{Results2}a. Once all accessible states in the monolayer are occupied, the Pauli exclusion principle blocks further transfer and the nanowire emission returns. This interpretation holds only if carrier diffusion and interlayer transport in the TMD take place on timescales comparable to the lasing dynamics of the nanowire.

\subsection{DFT Calculations}
\label{sec:dft}

Charge transfer across the interface between ZnO and a TMD monolayer such as MoS$_2$ or WS$_2$ is likely, and the literature reports both possible band alignments.\;In a type-I alignment,\cite{Bora.2024,Butanovs.2018} the conduction and valence band edges of the TMD both lie within the band gap of ZnO. After optical excitation, electrons and holes then both move from the nanowire into the monolayer, where they no longer contribute to lasing, and the nanowire is depleted of carriers.\;In a type-II alignment,\cite{Ali.2024,Jiang.2021,Jiang.2022,Liu.2016,Nguyen.2025,Nayak.2023} the carriers are spatially separated: electrons are transferred from the monolayer into the nanowire, whereas holes move from the nanowire into the monolayer. The studies differ mainly in which ZnO surface they consider and how clean it is.

To establish which of the two scenarios applies to our hybrid nanowires, we computed the band alignment between a monolayer of MoS$_2$ or WS$_2$ and the non-polar m-plane ZnO$(10\bar{1}0)$ facet, which is the surface contacted by the monolayer in the experiment, since the growth axis of the nanowires is along the c-axis. The band edges of ZnO are not reproduced by semilocal or standard hybrid functionals and the alignment is determined to a large extent by the electrostatic dipole that develops at the contact. We therefore evaluated the three contributions separately, each at the level of theory appropriate for it: the ionization potential (IP) and the electron affinity (EA) of the isolated, unstrained monolayers with HSE06 including spin-orbit coupling; the band edges of a symmetric eight-layer ZnO$(10\bar{1}0)$ slab with a screened hybrid functional containing a fraction equal to 0.375 of exact exchange, which reproduces the band gap and the Zn-3$d$ binding energy of ZnO; and the step $\Delta V$ of the plane-averaged electrostatic potential across the interface, obtained from supercells of 330 atoms in which the monolayer is strained by at most 1.2\%, relaxed with PBE+D3 and evaluated with the same hybrid functional as the ZnO slab. The band offsets then follow from $\Delta E_V = \mathrm{IP}^\mathrm{ZnO} - \mathrm{IP}^\mathrm{TMD} + \Delta V$ and $\Delta E_C = \mathrm{EA}^\mathrm{ZnO} - \mathrm{EA}^\mathrm{TMD} + \Delta V$, where positive values place the TMD edge above the ZnO edge.

\begin{figure}
    \centering
    \includegraphics[width=\linewidth]{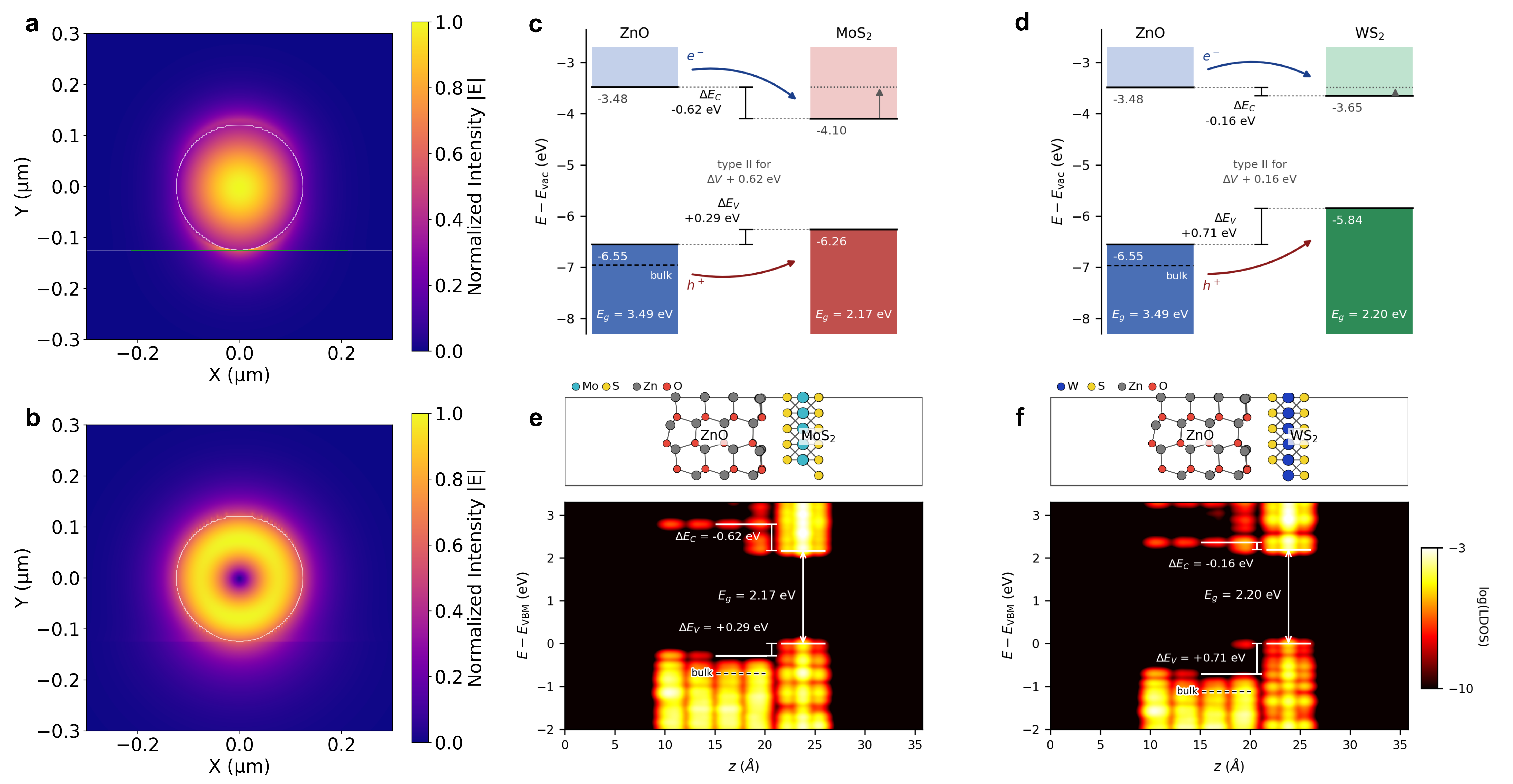}
    \caption{\justifying \textbf{Guided modes and calculated band alignment of monolayer TMD/ZnO heterojunctions.}
    (a) Electric field profiles of the fundamental HE$_{11}$ mode, showing a strongly confined and asymmetric field distribution and (b) TE$_{01}$ mode, characterized by a doughnut-shaped intensity profile.
    Calculated band alignment of (c) MoS$_2$/ZnO$(10\bar{1}0)$ and (d) WS$_2$/ZnO$(10\bar{1}0)$, relative to the vacuum level on the TMD side and including the interface dipole. Dark bars: valence bands; light bars: conduction bands; numbers: band-edge energies in eV. The dashed line in the ZnO valence band marks the bulk-like valence-band maximum, 0.41\,eV below the O-2$p$ surface state of the clean facet. At the ideal interface both photoexcited carriers are transferred into the monolayer (arrows). The dotted line and the grey arrow indicate the increase of $\Delta V$, that is, the reduction of the magnitude of the interface dipole, that turns the alignment into type II with electrons in ZnO, as predicted for a hydroxylated facet.
    Layer-resolved logarithm of the density of states across (e) the MoS$_2$/ZnO and (f) the WS$_2$/ZnO interface, with the relaxed structure drawn above on the same $z$ scale (half of the lateral supercell; Zn grey, O red, S yellow, Mo cyan, W blue). The eigenvalues of the 330-atom supercell are shifted, separately for each subsystem, to the band-edge positions of panels (c,d). White lines: valence and conduction band edges of each subsystem, taken as the extremal states carrying at least 30\% of their weight in that subsystem; dashed black line: bulk-like ZnO valence-band maximum. Energies refer to the valence-band maximum of the heterostructure.}

    \label{Results3}
\end{figure}

Upon relaxation, both monolayers approach the facet to 2.45\,\AA, with the lowest sulfur plane 2.63\,\AA\ (MoS$_2$) and 2.66\,\AA\ (WS$_2$) away from the threefold-coordinated zinc atoms of the surface dimers and with a flattening of the Zn-O dimers, which indicates a weak chemisorption rather than a van der Waals contact. The resulting interface dipole, $\Delta V = -0.82$\,eV for MoS$_2$/ZnO and $-0.81$\,eV for WS$_2$/ZnO, lowers the bands of the monolayer by about 0.8\,eV with respect to those of ZnO. Vacuum-level alignment alone would make both interfaces type II with electrons in ZnO. With the dipole included, the ideal interfaces are of type I, with the gap of the monolayer nested inside the gap of ZnO (\cref{Results3}c,d). The valence-band maximum lies in the monolayer, 0.29\,eV (MoS$_2$) and 0.71\,eV (WS$_2$) above the occupied O-2$p$ surface state that terminates the valence band of the clean facet, and 0.4 eV higher with respect to the bulk-like valence band of ZnO, while the conduction-band minimum of the monolayer lies 0.62\,eV (MoS$_2$) and 0.16\,eV (WS$_2$) below that of ZnO. The layer-resolved densities of states of the heterostructures in \cref{Results3}e,f show the same result in real space, with both band edges localized in the monolayer and a gap of 2.17\,eV (MoS$_2$/ZnO) and 2.20\,eV (WS$_2$/ZnO). For the pristine WS$_2$/m-plane-ZnO interface this alignment and a valence-band offset of about 0.6\,eV with the same ordering have been obtained independently in a recent hybrid-functional study.\cite{Sutarma.2026}

At the ideal, dry interface both photoexcited carriers are therefore transferred from the nanowire into the monolayer, where they no longer contribute to the gain. The type-I alignment accounts for the quenching of the emission and for the increase of the threshold without invoking band filling of the conduction band of ZnO. It also explains why the two hybrids behave identically, although their valence-band offsets differ by 0.42\,eV: both offsets exceed $k_\mathrm{B}T$ at room temperature by more than an order of magnitude, so that the hole transfer is in the saturated regime, in which the rate is governed by the interfacial coupling and by the density of accessible states of the monolayer and not by the size of the offset. The transfer is moreover self-limiting, because the holes accumulated in the monolayer lower its levels with respect to those of ZnO and reduce the offset that drives them: filling the valence band of MoS$_2$ over 0.29\,eV requires $7\times10^{13}$ holes per cm$^2$, which for the contact geometry of \cref{Results1}a corresponds to a few $10^{17}$ carriers per cm$^3$ in the nanowire, below the density reached at threshold, in agreement with the saturation of the charge transfer discussed in \cref{sec:photonic}. Since both carriers are captured, the capacity of the monolayer is set by the smaller of the two offsets, which is the valence offset for MoS$_2$ and the conduction offset of 0.16\,eV for WS$_2$. The two are of the same order, and so are the corresponding sheet densities, so that the same argument holds for both hybrids.

The conduction-band offsets are instead small, and $\Delta E_C$ depends linearly on the interface dipole: an increase of $\Delta V$ by 0.16\,eV for WS$_2$, or by 0.6\,eV for MoS$_2$, that is, a reduction of the magnitude of the dipole by the same amount, turns the alignment into type II with electron transfer into the nanowire. Changes of this size are expected at a facet that is neither ideal nor dry, and for a hydrated facet type II is indeed what is predicted: hydroxylation of the same m-plane surface has been shown to shift the conduction-band offset by about 1.3\,eV,\cite{Sutarma.2026} eight times more than WS$_2$ requires, and our nanowires are processed, transferred and measured in air. This reconciles our result with the photoemission data of Ref.~\cite{Eobaldt.2022}, which were deliberately recorded on transferred monolayers without any cleaning procedure, that is, on the hydrated surface, and which give type II with electron transfer into ZnO.
The two limits differ only in the fate of the electrons: the transfer of the holes into the monolayer, and with it the loss of one carrier species from the gain medium, is common to both, so that the suppression of the lasing does not depend on which of the two situations is realized at the facet.

The spectral position of the emission does not discriminate between the two alignments either. The hybridized nanowire lases at wavelengths at which the pristine nanowire shows no emission at all (\cref{Results1}e), so the gain spectrum itself is displaced, and the absorption of the monolayer, which can only remove modes from the envelope, cannot by itself produce the shift. The displacement follows from the higher carrier density at which the hybridized nanowire reaches threshold, and that increase is common to the two alignments, since in both of them the holes are extracted from the gain medium and the evanescent absorption is the same. Whether the displacement is dominated by the filling of the band edges or by the density dependence of the renormalized gap cannot be decided from these data. The wavelength dependence of the absorption accounts instead for the difference between the two hybrids: at the same high-energy cutoff, the MoS2-hybridized nanowire loses the low-energy modes that the WS2-hybridized one still supports (\cref{sec:lasing}). The alignment is therefore constrained by the quenching and by the threshold, which both limits reproduce, and not by the spectral position of the emission.

Either alignment removes carriers from the gain medium and thereby suppresses lasing. Because charge transfer depends strongly on carrier density, it can also explain why the influence of the hybridization grows with pump power. At low pump power density, the few carriers that leave the nanowire only slightly lower the luminescence intensity. In the  amplified spontaneous emission (ASE) and lasing regimes, charge transfer competes directly with stimulated emission, which accounts for the stronger suppression at high pump power densities. 
For semiconductor/TMD monolayer interfaces, charge transfer times from below 1\,ps to $\sim$\,20\,ps have been reported.\cite{Fang.2018} This is considerably faster than the radiative lifetimes of ZnO, which range from hundreds of ps to ns,\cite{Ali.2024} 
 and faster than the lasing dynamics, so charge transfer can compete with stimulated emission and outpace spontaneous recombination. For pristine ZnO nanowires with diameters above 200\,nm, lasing sets in after about 1\,ps and lasts 5-20\,ps (pulse width).\cite{Sidiropoulos.2014,Roder.2015}

\section{Conclusion}

We examined loss mechanisms in ZnO nanowire hybridized with WS$_2$ and MoS$_2$ monolayer, by obtaining lasing thresholds. Hybridization reduces the threshold with increasing overlap between the hybridization partners.
Two independent mechanisms were identified by combining further experiments and numerical calculations.
A monolayer adds a modal loss of 1000-1150\,cm$^{-1}$ to the TE$_{01}$ mode, a large part of the available gain of 1600-2800\,cm$^{-1}$.\;From the bilayer on, the loss reaches 1850\,cm$^{-1}$ and more, comparable to the available gain, and lasing disappears. 
Because this loss depends on wavelength, it removes modes from the low-energy side of the envelope and accounts for the 0.4\,nm by which the shifts of the two monolayers differ, while the blueshift itself follows the increase of the threshold, at which the hybridized nanowire lases outside the emission window of the pristine one. 
Second, the m-plane facet of the nanowire and the monolayer form a type-I junction when the interface is ideal, mostly unstrained and dry. 
Both photoexcited carriers then move into the monolayer, which explains the quenched emission and the more than doubled threshold. 
The threshold is the same for both TMDs because the hole transfer is saturated. 
The conduction-band offsets are small (0.16\,eV for WS$_2$, 0.62\,eV for MoS$_2$), so the direction of the electron transfer depends on the chemical state of the facet, and a hydroxylated, air-exposed facet is expected to give type II.
In both cases, the holes leave the nanowire for the monolayer, and lasing is suppressed.

\section*{Author Contributions}
\textbf{M.W.} and \textbf{E.E.} contributed equally to this work.
\textbf{M.W.}: Formal analysis, Visualization, Writing - original draft.
\textbf{E.E.}: Investigation, Methodology, Visualization, Formal analysis.
\textbf{A.Z.}, \textbf{A.E.}: Investigation.
\textbf{M.T.}: Resources (TMD crystals for exfoliation).
\textbf{A.G.}: Resources (growth of monolayer WS$_2$ sample).
\textbf{L.R.J.}: Investigation (FDE simulations).
\textbf{X.C.}, \textbf{P.S.}, \textbf{N.T.-D.}: Investigation, Formal analysis (DFT calculations).
\textbf{A.T.}, \textbf{S.B.}, \textbf{G.S.}: Supervision.
\textbf{C.R.}: Conceptualization, Supervision, Funding acquisition, Writing - review \& editing.
All authors discussed the results and commented on the manuscript.

\section{Acknowledgments}

The authors acknowledge the German Research Foundation (Deutsche Forschungsgemeinschaft) DFG via the SFB 1375 ``NOA - Non-linear optics down to atomic scales'' projects A2, B2 and B5 (project number 398816777). 
During the preparation of this work, the authors used Claude Opus 5.0 to support language editing. After using this tool, the authors reviewed and edited the content as needed and take full responsibility for the content of the publication.

\section{Experimental Section and Simulation Parameters}

\subsection{ZnO Nanowire Growth and TMD Monolayer Preparation}
ZnO nanowires (NWs) were synthesized via vapor transport using the vapor-liquid-solid (VLS) mechanism.\cite{Wagner.1964}
A ZnO/graphite (7:1) precursor was evaporated at 1050\,°C in a horizontal tube furnace.\cite{Borchers.2006} An Ar/O$_2$ flow (10\,sccm each) then transported the vapor to Si substrates coated with a 10\,nm Au film. The Au forms catalytic droplets that define the NW diameter, whereas the length is controlled by growth time and vapor pressure. Growth was performed at 7\,mbar for 1\,h, yielding NWs with diameters of 150-300\,nm and lengths of 10-25\,µm.\cite{Eobaldt.2026}

TMD monolayers were prepared either by mechanical exfoliation or by chemical vapor deposition (CVD). Exfoliated flakes with lateral sizes of 10-100\,µm were identified by their optical contrast and transferred onto SiO$_2$/Si substrates.\cite{Splendiani.2010} Alternatively, WS$_2$ monolayers were grown by CVD from WO$_3$ and elemental sulfur precursors in a two-zone furnace, which yields triangular domains with side lengths of 10-50\,µm.\cite{George.2019,Shree.2020} The growth is described in detail in Ref.~\cite{George.2019}.

\subsection{Nanowire Transfer and Nanomanipulation}

Nanowires were positioned with two complementary techniques.
For deterministic placement, we used a stage-assisted transfer with a polydimethylsiloxane (PDMS) stamp: as-grown nanowires were first dry-imprinted onto a SiO$_2$/Si substrate, and individual wires were then picked up with the stamp and released onto the target substrate in a commercial transfer stage (HQ Graphene).
Where higher precision was required, nanowires were moved by tip-assisted nanomanipulation. The nanomanipulator (MM3A-EM, Kleindiek Nanotechnik) carries a $\sim$1~$\mu$m tungsten tip that is moved along all three spatial directions by piezoelectric motors. Further details are given in Ref.~\cite{Eobaldt.2026}.

\subsection{Micro-Photoluminescence ($\mu$PL) Measurements}

The optical characterization of single nanowires was performed using a micro-photoluminescence ($\mu$PL) setup at room temperature under ambient conditions.
Optical excitation was provided by a frequency-tripled Nd:YAG laser operating at 355\,nm with a 7\,ns pulse duration and a 100\,Hz repetition rate. The beam was focused onto the sample by a 50$\times$ near-UV microscope objective (NA\,=\,0.43), which also collected the photoluminescence. A diverging lens in the excitation path widened the spot to 20-25\,µm, so that the whole nanowire was illuminated. The emission was spectrally filtered using a long-pass filter and analyzed by a 500\,mm monochromator (Princeton Instruments SP-2500i) with a grating with 1200\,grooves/mm. The sample position could be adjusted with nanometer precision using a piezo-controlled stage to ensure optimal alignment of individual nanowires within the excitation spot.

\subsection{Finite-Difference Eigenmode Simulations}

The FDE simulations were carried out with Lumerical MODE Solutions at a wavelength of 385\,nm. ZnO was described by $n = 2.25$, which corresponds to the high-excitation regime,\cite{Wille.2016} and $k = 0$, i.e., no absorption inside the nanowire. For SiO$_2$ we used $n = 1.47$ and $k = 0$,\cite{Arosa.2020} for WS$_2$ $n = 2.74$ and $k = 2.80$,\cite{Ermolaev.2020} and for the MoS$_2$ monolayer $n = 2.18$ and $k = 3.27$.\cite{Song.2019}

\subsection{Density Functional Theory}

Density-functional calculations were performed with VASP\cite{Kresse.1996} using projector-augmented-wave potentials,\cite{Blochl.1994,Kresse.1999} a plane-wave cutoff of 500\,eV and experimental lattice parameters (ZnO $a = 3.2496$\,\AA, $c = 5.2042$\,\AA, $u = 0.3819$; MoS$_2$ $a = 3.160$\,\AA; WS$_2$ $a = 3.154$\,\AA). All band edges were referred to the vacuum level obtained from the plane-averaged electrostatic potential. The isolated monolayers (PBE-relaxed positions, $12\times12$ $k$ points, 20\,\AA\ of vacuum) were computed with HSE06.\cite{Heyd.2003,Heyd.2006,Krukau.2006} Spin-orbit coupling was included as a valley-resolved shift of the band edges obtained from PBE and PBE+SOC calculations. The ZnO$(10\bar{1}0)$ surface was modeled by a symmetric slab of eight Zn$_2$O$_2$ layers with PBE-relaxed positions and 20\,\AA\ of vacuum, and its band edges were computed with a screened hybrid functional with 37.5\% of exact exchange and $\mu = 0.2$\,\AA$^{-1}$, which reproduces the band gap of ZnO.\cite{Oba.2008} The interface was modeled by a $5\times3$ surface supercell of ZnO$(10\bar{1}0)$ with four layers (240 atoms) and a $3\times5$ supercell of the rectangular TMD cell (90 atoms), in which the monolayer is strained by at most 1.2\%. All atoms were relaxed with PBE\cite{Perdew.1996} and the D3(BJ) dispersion correction\cite{Grimme.2010,Grimme.2011} to forces below 0.02\,eV/\AA. The potential step across the interface was obtained from static calculations with a dipole correction along the surface normal,\cite{Neugebauer.1992} both with PBE+D3 and with the 37.5\% hybrid functional at the same geometry. The band offsets follow from $\Delta E_V = \mathrm{IP}^\mathrm{ZnO} - \mathrm{IP}^\mathrm{TMD} + \Delta V$ and $\Delta E_C = \mathrm{EA}^\mathrm{ZnO} - \mathrm{EA}^\mathrm{TMD} + \Delta V$.\cite{Anderson.1962,Kang.2013,VandeWalle.1987} Convergence tests, the comparison of exchange-correlation functionals, direct hybrid-functional calculations of the 330-atom heterostructures and the error budget of the offsets are reported in the Supporting Information.

\printbibliography

\end{document}

\begin{center}
\textbf{\large Supporting Information}\\[1ex]
\textbf{Photonic and Electronic Interaction of TMD Hybridized ZnO Nanowire Lasers}\\[1ex]
Moritz Willems*$^a$, Edwin Eobaldt*$^a$, Alexander Zaunick$^a$, Aurelia Eberhard$^a$, Maximilian Tomoscheit$^a$, Lukas Raam Jäger$^a$, Xiao Chen$^b$, Preeti Sharma$^b$, Nicolas Tancogne-Dejean$^b$, Antony George$^c$, Andrey Turchanin$^{c,d}$, Silvana Botti$^b$, Giancarlo Soavi$^{a,d}$ and Carsten Ronning$^{a,d}$
\newline
\end{center}
\noindent
$^a$Institute of Solid State Physics, Friedrich Schiller University Jena, 07743 Jena, Germany
\newline
$^b$RC Future Energy Materials and Systems $\&$ ICAMS, Faculty of Physics, Ruhr University Bochum, 44801 Bochum, Germany
\newline
$^c$Institute of Physical Chemistry, Friedrich Schiller University Jena, 07743 Jena, Germany
\newline
$^d$Abbe Center of Photonics, Friedrich Schiller University Jena, 07745 Jena, Germany

\section{Raman and PL identification of the MoS$_2$ layer number}

\begin{figure}[H]
\centering
\includegraphics[width=\linewidth]{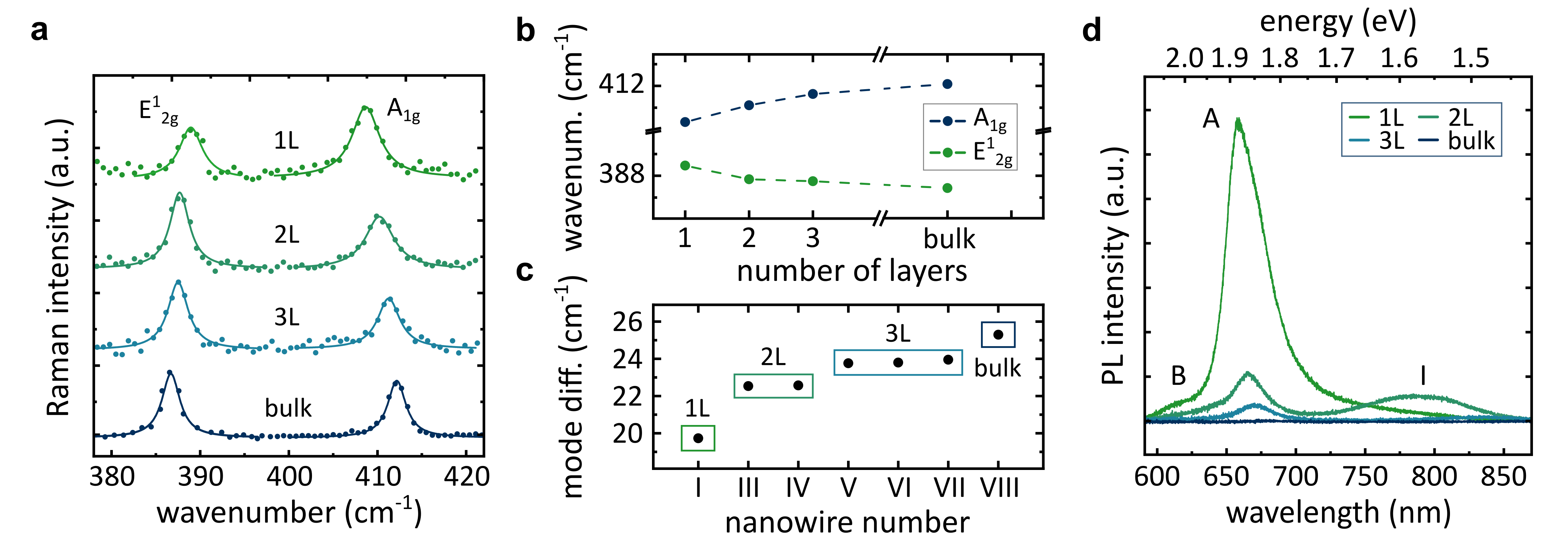}
\caption{\justifying \textbf{Layer-number identification of the exfoliated MoS$_2$ flakes.}
(a) Raman spectra of a MoS$_2$ monolayer (1L, green), bilayer (2L, cyan), trilayer (3L, light blue)
and bulk flake (dark blue) at 532\,nm excitation; solid lines are Gaussian fits of the
$E^1_{2g}$ and $A_{1g}$ modes.
(b) Peak positions of the two modes from the fits in (a) versus layer number.
(c) Difference between the $A_{1g}$ and $E^1_{2g}$ frequencies for the flakes of
nanowires I-VIII, grouped by their layer number.
(d) PL spectra of of MoS$_2$ flakes with four different layer numbers, with the A and B excitons and the indirect-gap
emission (I).}
\label{RamanNW}
\end{figure}

The MoS$_2$ flakes for the layer-dependent experiments (Sec. 2.2 of the main text) were exfoliated and preselected by their optical contrast in the transfer-stage microscope.\cite{Bing.2018} After transfer onto a clean SiO$_2$/Si substrate, we determined the layer number of each flake by Raman and PL spectroscopy at room temperature. A 532\,nm continuous-wave laser (Cobolt 08-DPL) was focused to a spot of about 1\,\textmu m by a 50$\times$ objective (NA = 0.42), and the signal was analyzed by a spectrometer (Horiba iHR 550) with an 1800\,grooves/mm grating at a resolution of about 0.7\,cm$^{-1}$.

Fig.~S1a shows the two Raman modes of MoS$_2$. The lower-frequency mode is the in-plane $E^1_{2g}$ mode, in which Mo
and S atoms vibrate against each other, and the higher-frequency mode is the out-of-plane $A_{1g}$ mode of the sulfur
atoms. As the layer number increases, van der Waals coupling between the layers stiffens the
$A_{1g}$ mode and dielectric screening softens the $E^1_{2g}$ mode, so the two modes move
apart (Fig.~S1b).\cite{Lee.2010} The flakes of nanowires I-VIII fall into four groups with
mode differences of 19.7, 22.5, 23.7 and 25.3\,cm$^{-1}$ with an uncertainty of $\pm$0.1\,cm$^{-1}$ (Fig.~S1c),
which we assign to a monolayer, bilayer, trilayer and bulk MoS$_2$.\cite{Lee.2010} Raman
spectroscopy cannot give the layer number of the bulk flake, because the difference hardly
changes for thick flakes. Its optical contrast indicates at least 100 layers.

The PL spectra (Fig.~S1d) support this assignment. The A and B excitons come from direct
transitions at the K point and redshift and weaken with increasing layer
number. Only the monolayer has a direct band gap, so it emits
much more strongly than the thicker flakes.\cite{Mak.2010, Splendiani.2010} The multilayers
also show weak emission from the indirect gap (I). Its redshift with layer
number cannot be resolved at this intensity.\cite{Mak.2010}

\section*{\texorpdfstring{Band alignment of monolayer MoS$_2$ and WS$_2$ on the non-polar m-plane \ZnOm\ facet}{Band alignment of monolayer MoS2 and WS2 on the non-polar m-plane ZnO facet}}

\section{Computational details}
\label{sec:details}

All calculations were performed with VASP 6.6.1 \cite{Kresse.1996} and projector-augmented-wave potentials \cite{Blochl.1994,Kresse.1999} with the valence configurations Zn ($3d^{10}4s^{2}$), O ($2s^{2}2p^{4}$), Mo ($4p^{6}4d^{5}5s^{1}$), W ($5s^{2}5p^{6}5d^{4}6s^{2}$) and S ($3s^{2}3p^{4}$). Common settings were a plane-wave cutoff of 500\,eV, \texttt{PREC = Accurate}, Gaussian smearing of 0.02\,eV, an aspherical treatment of the gradient corrections inside the PAW spheres, and an electronic convergence of $10^{-6}$\,eV. The hybrid calculations use a screening parameter of 0.2\,\AA$^{-1}$ and the all-band conjugate-gradient algorithm, started from converged PBE wave functions. Vacuum levels were taken from the plane-averaged Hartree-plus-ionic potential, averaged over a 2\,\AA\ window centered 6\,\AA\ outside the outermost atoms. In every calculation the potential is flat within that window to better than 0.1\,meV. Ionization potentials and electron affinities were obtained from the highest occupied and the lowest unoccupied eigenvalue over the full $k$-point mesh. We use experimental lattice parameters throughout: ZnO $a = 3.2496$\,\AA, $c = 5.2042$\,\AA, $u = 0.3819$; MoS$_2$ $a = 3.160$\,\AA; WS$_2$ $a = 3.154$\,\AA. The ZnO lattice defines the in-plane periodicity of the interface, because the nanowire is a bulk crystal, and the monolayer is strained to match it.

\begin{table}[H]
\centering
\caption{\justifying Calculations and parameters. ``SOC corr.'' denotes the valley-resolved PBE$\to$PBE+SOC shift of the band edges.}
\label{tab:S1}
\footnotesize
\setlength{\tabcolsep}{2.5pt}
\begin{tabular}{lllll}
\toprule
system & functional & $k$ mesh & relaxation & purpose\\
\midrule
MoS$_2$, WS$_2$ 1L & PBE & $12\times12$ & ions, $<0.01$\,eV/\AA & geometry at experimental $a$\\
 & PBE+SOC & $12\times12$ & -- & SOC corr.\\
 & HSE06, HSE(0.375) & $12\times12$ & -- & IP, EA\\
 & PBE, strained cell & $8\times14$ & uMLIP & strain test\\
\ZnOm\ 8L, 4L & PBE & $8\times5$ & ions, $<0.01$\,eV/\AA & geometry, thickness test\\
\ZnOm\ 8L & HSE(0.375), HSE06 & $6\times4$ & -- & IP, EA\\
interface, 330 atoms & PBE+D3 & $\Gamma$ & all ions, $<0.02$\,eV/\AA & geometry\\
 & PBE+D3, dipole corr. & $\Gamma$, $2\times2$ & -- & $\Delta V$, $k$ convergence\\
 & HSE(0.375), HSE06, dipole corr. & $\Gamma$ & -- & hybrid $\Delta V$, direct band edges\\
 & PBE+D3, non-self-consistent & $(0,\tfrac{1}{3},0)$ & -- & K valley at the interface\\
1L, rectangular cell & PBE, hybrids & $3\times5$, $6\times15$ & ions (PBE) & K-valley corr.\\
\bottomrule
\end{tabular}
\end{table}

The interface cell is asymmetric, so the final static calculations include a dipole correction along the surface normal \cite{Neugebauer.1992}, with the dipole layer at the center of the vacuum region. This makes the potential flat in both vacuum regions and gives the potential step directly. The relaxations were performed without the dipole correction, which with 20\,\AA\ of vacuum has a negligible effect on the forces. The isolated ZnO slabs and the strained monolayer fragments were pre-relaxed with a universal machine-learning interatomic potential trained on the MatPES PBE dataset \cite{Kaplan.2025}, and all geometries reported here are fully relaxed with DFT.

\section{The three-step alignment scheme}
\label{sec:scheme}

A direct hybrid-functional calculation of the heterostructure requires either very large cells or small commensurate cells with a strain of several percent in the monolayer, which shifts the TMD band edges by several tenths of an eV and can change the character of the gap. We therefore use the standard decomposition of a band offset into the band-edge positions of the separated components and an interface potential step \cite{Anderson.1962,Kang.2013,VandeWalle.1987, Hinuma.2013}, in which each term can be computed at the level of theory appropriate for it,
\begin{align}
\Delta E_V &= E_V^\mathrm{TMD} - E_V^\mathrm{ZnO} = \mathrm{IP}^\mathrm{ZnO} - \mathrm{IP}^\mathrm{TMD} + \Delta V, \label{eq:dEV}\\
\Delta E_C &= E_C^\mathrm{TMD} - E_C^\mathrm{ZnO} = \mathrm{EA}^\mathrm{ZnO} - \mathrm{EA}^\mathrm{TMD} + \Delta V, \label{eq:dEC}\\
\Delta V &= V_\mathrm{vac}^\mathrm{TMD\ side} - V_\mathrm{vac}^\mathrm{ZnO\ side}. \label{eq:dV}
\end{align}
IP and EA are the ionization potential and the electron affinity of the isolated systems, that is, the valence-band maximum and the conduction-band minimum measured from their own vacuum level. $\Delta V$ is the difference between the plane-averaged electrostatic potentials in the two vacuum regions of the dipole-corrected heterostructure and contains all interface effects: charge transfer, Pauli push-back, the change of the ZnO surface dipole upon contact and the structural relaxation of both surfaces. The same decomposition is usually applied with the macroscopic average of the electrostatic potential inside each material, or with core levels, as the reference \cite{VandeWalle.1987, Hinuma.2013}; here the supercell has vacuum on both sides, so that the vacuum level can be used directly, which has the further advantage of giving the absolute ionization potentials. For $\Delta V = 0$ the equations reduce to Anderson's electron-affinity rule.\cite{Anderson.1962, Oba.2018} $\Delta V$ is a ground-state electrostatic quantity and is computed in the large, almost unstrained supercells. IP and EA, which depend strongly on the functional, are computed with hybrid functionals for the small isolated systems.

\textbf{Choice of functionals.} For ZnO, HSE06 ($\alpha = 0.25$) gives a band gap of about 2.5\,eV, while the larger fraction $\alpha = 0.375$ with $\mu = 0.2$\,\AA$^{-1}$ \cite{Oba.2008} reproduces the experimental gap and the position of the Zn-$3d$ states. The same strategy of adjusting the fraction of exact exchange to the measured gap of the oxide has been followed for other oxide interfaces, for instance for strontium germanate on silicon.\cite{Rauch.2023} We use the modified HSE06 for ZnO and for the interface dipole, reporting HSE06 for comparison. For the monolayers, HSE06 with spin-orbit coupling gives gaps of 2.17 and 2.20\,eV, whereas $\alpha = 0.375$ opens them by a further 0.25\,eV. We use HSE06+SOC as the reference and report both. Spin-orbit coupling is included for the TMDs as the correction $\delta^\mathrm{SOC}(\mathbf{k}) = \varepsilon^\mathrm{PBE+SOC}(\mathbf{k}) - \varepsilon^\mathrm{PBE}(\mathbf{k})$ of the band-edge states, applied valley by valley (valence band at K and $\Gamma$, conduction band at K and Q) before the band extrema are taken: the spin-orbit splittings are almost independent of the functional, but the relative energies of the K and Q valleys are not, so a global shift of the extrema would be wrong for WS$_2$. For ZnO the spin-orbit splittings are of the order of 10\,meV and are neglected. The interface geometry and $\Delta V$ use PBE \cite{Perdew.1996} with Grimme's D3 dispersion correction and Becke-Johnson damping.\cite{Grimme.2010,Grimme.2011}

\section{Isolated monolayers}
\label{sec:monolayers}

\begin{table}[htbp]
\centering
\caption{\justifying Band edges of the isolated, unstrained monolayers relative to the vacuum level (eV). The ``+SOC'' rows contain the valley-resolved PBE spin-orbit correction; the shifts $\delta$IP and $\delta$EA are given in the PBE+SOC row. The last row of each block is the monolayer strained as in the interface supercell. Bold: values used for the final offsets.}
\label{tab:S2}
\vspace{0.3cm}
\begin{tabular}{llccc}
\toprule
 & functional & IP & EA & $E_g$\\
\midrule
MoS$_2$ & PBE & 5.89 & 4.13 & 1.76\\
 & PBE+SOC & 5.82 ($-0.07$) & 4.14 ($+0.01$) & 1.68\\
 & HSE06 & 6.33 & 4.09 & 2.25\\
 & HSE(0.375) & 6.55 & 4.07 & 2.47\\
 & \textbf{HSE06+SOC} & \textbf{6.26} & \textbf{4.10} & \textbf{2.17}\\
 & HSE(0.375)+SOC & 6.47 & 4.08 & 2.40\\
 & PBE, strained as in the interface & 5.86 & 3.98 & 1.88\\
\midrule
WS$_2$ & PBE & 5.64 & 3.68 & 1.96\\
 & PBE+SOC & 5.43 ($-0.21$) & 3.83 ($+0.15$) & 1.60\\
 & HSE06 & 6.05 & 3.58 & 2.47\\
 & HSE(0.375) & 6.26 & 3.53 & 2.73\\
 & \textbf{HSE06+SOC} & \textbf{5.84} & \textbf{3.65} & \textbf{2.20}\\
 & HSE(0.375)+SOC & 6.05 & 3.60 & 2.45\\
 & PBE, strained as in the interface & 5.62 & 3.68 & 1.94\\
\bottomrule
\end{tabular}
\end{table}

Increasing $\alpha$ from 0.25 to 0.375 lowers the valence-band maximum of the monolayers by about 0.2\,eV and leaves the conduction-band minimum almost unchanged. Spin-orbit coupling raises the valence-band maximum at K by half of the spin-orbit splitting of the valence band. The conduction band of WS$_2$ requires care: without spin-orbit coupling the conduction-band minimum is at K in PBE and in both hybrids, whereas spin-orbit coupling lowers the Q valley, halfway along $\Gamma$-K, more than K, so that in PBE+SOC the minimum moves to Q, 84\,meV below K. Applying the PBE shift of the global minimum to the hybrid K-valley minimum would overestimate the electron affinity by 0.08\,eV; with the valley-resolved correction the K and Q valleys of WS$_2$ are degenerate within 3\,meV in HSE06+SOC. The compressive strain of about 1\% present in the interface supercell changes the PBE band edges by at most 0.15\,eV, mostly in the conduction band. We use the band edges of the unstrained monolayers, which is the situation of the experiment, where the monolayer is not lattice-matched to the nanowire. The residual strain of the supercell enters only through $\Delta V$, to which it contributes less than 0.05\,eV.

\begin{figure}[H]
\centering
\includegraphics[width=\linewidth]{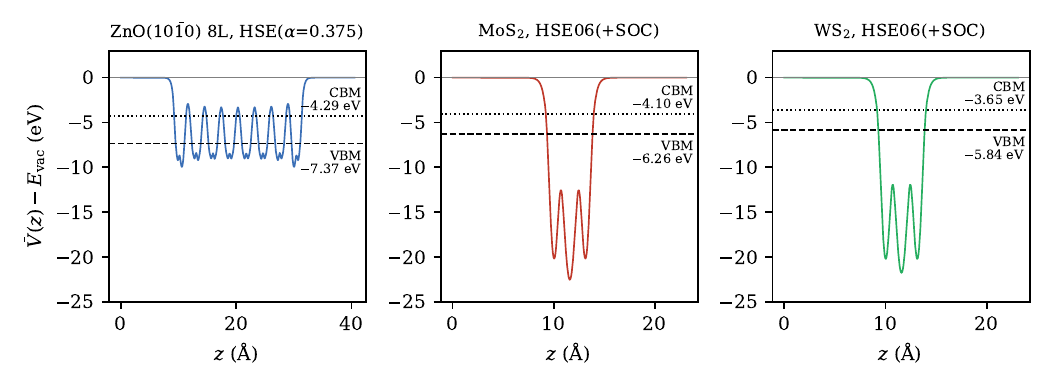}
\caption{\justifying Plane-averaged electrostatic potential of the isolated systems relative to their own vacuum level, with the valence-band maximum (dashed) and the conduction-band minimum (dotted): eight-layer \ZnOm\ slab with HSE(0.375), MoS$_2$ and WS$_2$ with HSE06 including the spin-orbit correction.}
\label{fig:S1}
\end{figure}

\section{\texorpdfstring{The \ZnOm\ slab and its O-$2p$ surface state}{The ZnO(10-10) slab and its O-2p surface state}}
\label{sec:zno}

\begin{table}[H]
\centering
\caption{\justifying Band edges of the symmetric \ZnOm\ slabs relative to vacuum (eV). IP$_\mathrm{surf}$ refers to the slab valence-band maximum, which is a surface state, IP$_\mathrm{bulk}$ to the highest occupied state of bulk character. The gaps are $E_g = \mathrm{IP} - \mathrm{EA}$.}
\label{tab:S3}
\vspace{0.3cm}
\begin{tabular}{lccccc}
\toprule
 & IP$_\mathrm{surf}$ & IP$_\mathrm{bulk}$ & EA & $E_g$(surf) & $E_g$(bulk)\\
\midrule
PBE, 4 layers & 5.82 & -- & 4.95 & 0.87 & --\\
PBE, 8 layers & 5.81 & 6.10 & 5.14 & 0.67 & 0.96\\
HSE06, 8 layers & 6.83 & 7.20 & 4.60 & 2.22 & 2.60\\
\textbf{HSE(0.375), 8 layers} & \textbf{7.37} & \textbf{7.78} & \textbf{4.29} & \textbf{3.07} & \textbf{3.49}\\
\bottomrule
\end{tabular}
\end{table}

The ionization potential is converged with respect to the slab thickness (5\,meV between 4 and 8 layers in PBE), whereas the electron affinity of the 4-layer slab is 0.19\,eV smaller because of the quantum confinement of the dispersive Zn-$4s$ conduction band. We therefore use the 8-layer values. Projecting the slab states on the outermost Zn$_2$O$_2$ unit of each side, 25\% of the atoms, shows that the two highest occupied states, one per surface, carry 89\% of their weight on those atoms, whereas the conduction-band minimum carries only 14\%. The valence-band maximum of the clean facet is thus the occupied O-$2p$ dangling-bond state of the surface dimers, 0.41\,eV above the highest bulk-like valence state with HSE(0.375) and 0.28\,eV in PBE. With respect to the bulk-like valence band the HSE(0.375) gap is 3.49\,eV, in agreement with the experimental gap of 3.44\,eV. The smaller slab gap of 3.07\,eV is entirely due to the surface state. Both energies are relevant for a nanowire facet: the surface state is the topmost occupied level of the clean facet, and holes in the interior of the nanowire sit at the bulk-like valence-band maximum. The bulk-like ionization potential, 7.78\,eV, agrees with the measured value of 7.8\,eV for the non-polar $(10\bar{1}0)$ and the O-terminated $(000\bar{1})$ surfaces of ZnO, whereas the Zn-terminated $(0001)$ surface has an ionization potential of 6.9\,eV.\cite{Klein.2010} This difference of 0.9\,eV between the polar c-plane and the non-polar m-plane surfaces is the reason why photoemission data measured on c-plane crystals cannot be transferred to the side facets of the nanowires.

In PBE the ionization potential of the ZnO slab is 5.81\,eV, 1.55\,eV above the HSE(0.375) value, and the slab gap is only 0.67\,eV. The PBE valence bands of ZnO and of the monolayers then lie at practically the same energy (5.81 against 5.89 and 5.64\,eV), so that small interface effects decide which of the two is higher. This is the origin of the valence-band ordering obtained with semilocal functionals, discussed in \cref{sec:offsets} and \cref{sec:ldos}.

\begin{figure}[H]
\centering
\includegraphics[width=0.85\linewidth]{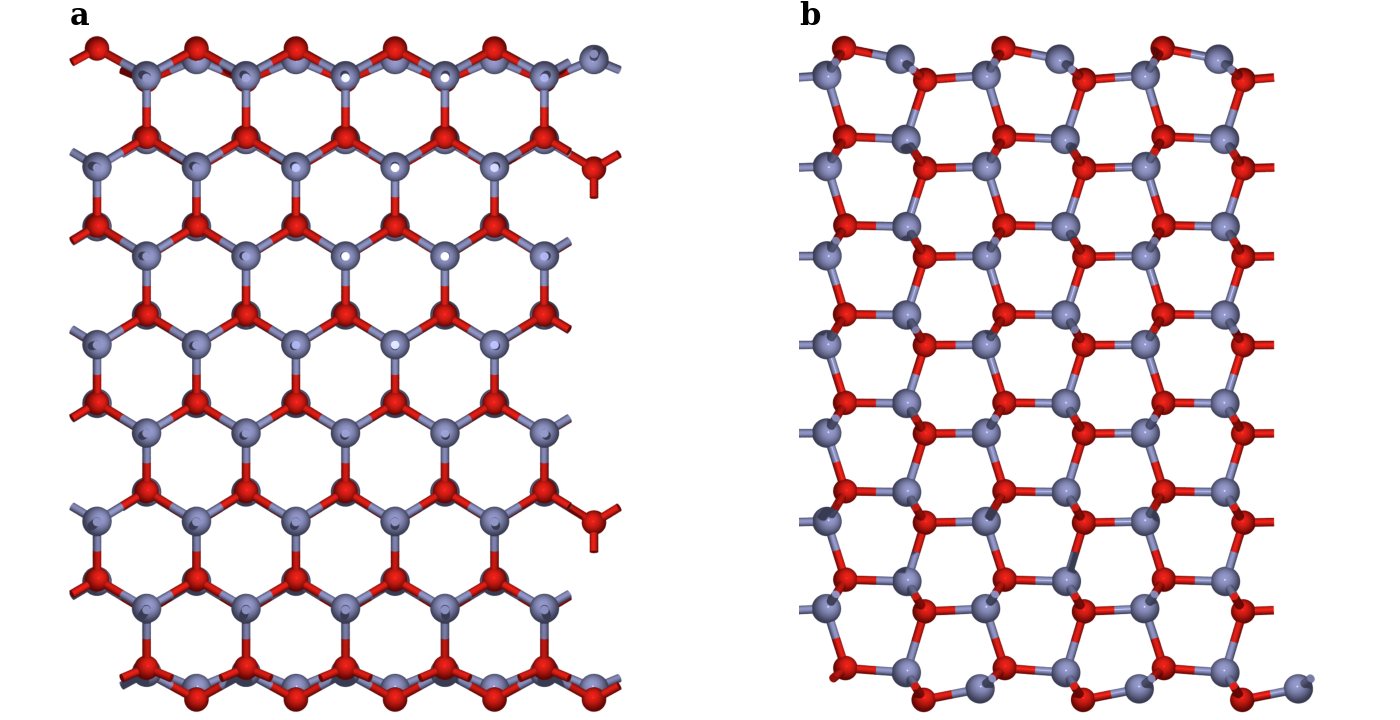}
\caption{\justifying Symmetric eight-layer \ZnOm\ slab with PBE-relaxed positions, repeated laterally for clarity (Zn gray-blue, O red). (a) View along the $c$ axis. (b) View along the $a$ axis, showing the Zn-O dimers of the two surfaces, with the oxygen displaced outwards by 0.62\,\AA.}
\label{fig:S2}
\end{figure}

\section{Interface model and geometry}
\label{sec:geometry}

The interface cell consists of $5\times3$ \ZnOm\ surface cells ($5a\times3c = 16.248\times15.613$\,\AA$^2$) with four ZnO layers (240 atoms, one Zn$_2$O$_2$ double plane per layer) and a $3\times5$ supercell of the rectangular ($\sqrt{3}a\times a$) TMD cell (90 atoms) rotated so that the zigzag direction of the monolayer lies along ZnO $[0001]$. The 330 atoms carry about 2900 valence electrons. The resulting strains of the monolayer are $-1.05\%/-1.19\%$ for MoS$_2$ and $-0.86\%/-1.00\%$ for WS$_2$ along the two in-plane directions, an order of magnitude smaller than in the 30-atom commensurate cells that are often used for this interface. The monolayer was placed with its lowest sulfur plane 2.9\,\AA\ above the topmost ZnO atom and all atoms were relaxed.

\begin{table}[H]
\centering
\caption{\justifying Geometry of the relaxed PBE+D3 interfaces (\AA). $d_\perp$ is the vertical distance between the lowest S plane and the topmost ZnO atom, $\Delta z_\mathrm{O-Zn}$ is the outward displacement of the surface O with respect to the surface Zn of the dimers, which is 0.62\,\AA\ at the clean surface.}
\label{tab:S4}
\vspace{0.3cm}
\begin{tabular}{lcccccc}
\toprule
 & $d_\perp$ & $d_\perp$ initial & min $d_\mathrm{S-Zn}$ & min $d_\mathrm{S-O}$ & $\Delta z_\mathrm{O-Zn}$ & ionic steps, $F_\mathrm{max}$ (eV/\AA)\\
\midrule
MoS$_2$/ZnO & 2.45 & 2.90 & 2.63 & 2.68 & 0.28 & 66, 0.018\\
WS$_2$/ZnO & 2.45 & 2.90 & 2.66 & 2.70 & 0.26 & 57, 0.015\\
\bottomrule
\end{tabular}
\end{table}

\begin{figure}[H]
\centering
\includegraphics[width=\linewidth]{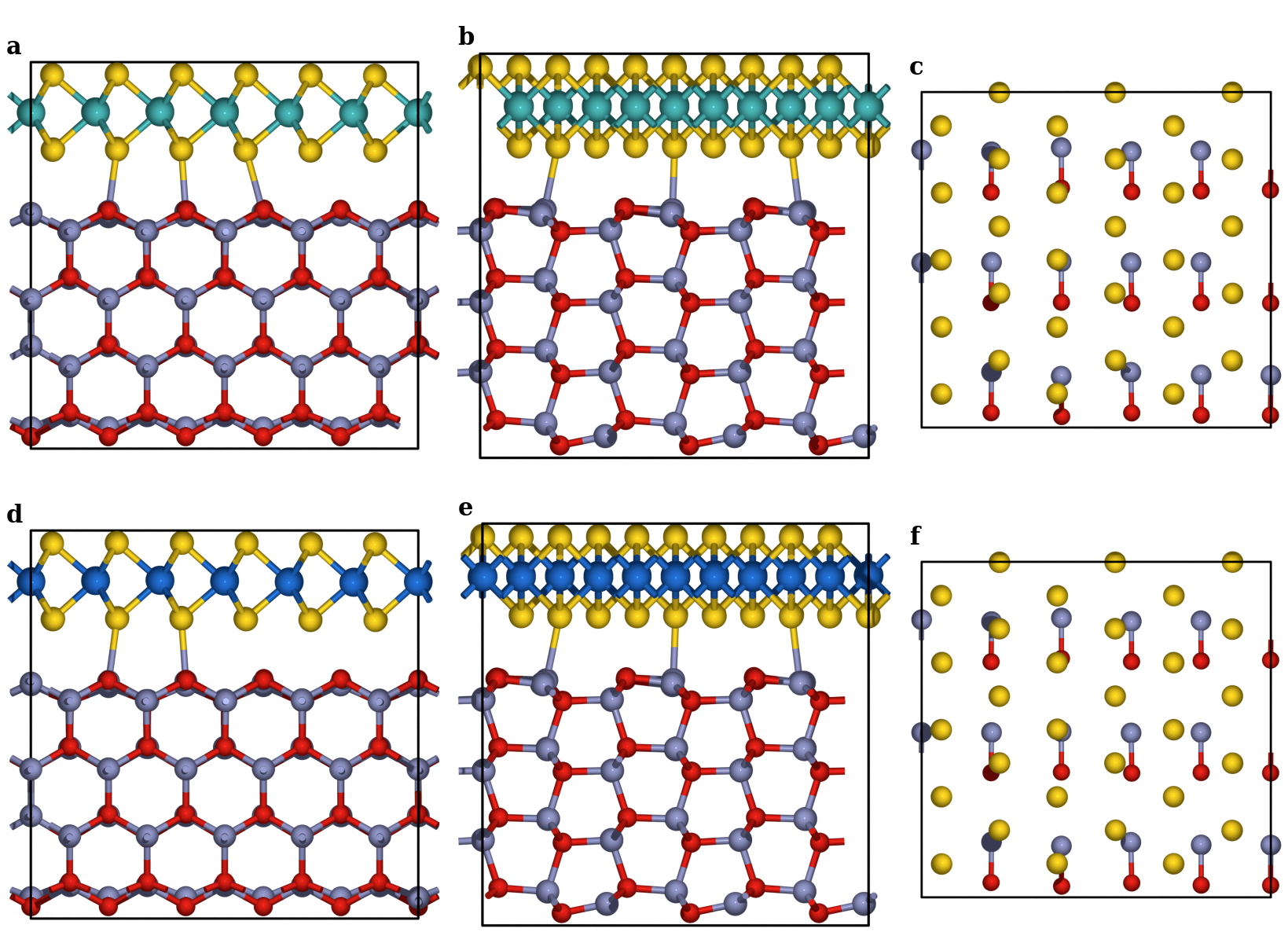}
\caption{\justifying Relaxed PBE+D3 structures of the 330-atom interface cells (Zn gray-blue, O red, S yellow, Mo teal, W blue; the frame is the $16.25\times15.61$\,\AA$^2$ supercell). (a,b,d,e) Side views along the two in-plane axes for MoS$_2$/ZnO (a,b) and WS$_2$/ZnO (d,e). Bonds are drawn for Zn-O $<2.15$\,\AA, metal-S $<2.6$\,\AA\ and for the interfacial S-Zn contacts $<2.8$\,\AA. (c,f) Topmost Zn-O dimer layer and the lowest S plane seen from above, showing the incommensurate registry.}
\label{fig:S3}
\end{figure}

Both monolayers approach the surface from the initial 2.9\,\AA\ to $d_\perp \approx 2.45$\,\AA, with S-Zn distances of 2.63-2.66\,\AA. These distances are longer than the Zn-S bond in ZnS (2.34\,\AA) but much shorter than a van der Waals contact (about 3.2\,\AA), and the surface dimers flatten, the outward tilt of the oxygen decreasing from 0.62 to about 0.27\,\AA\ as the threefold-coordinated Zn atoms move towards the sulfur plane. The interaction is therefore a weak chemisorption of the monolayer on the coordinatively unsaturated surface Zn atoms, in agreement with the value of 2.615\,\AA\ reported for the pristine WS$_2$/\ZnOm\ interface.\cite{Sutarma.2026} Only a fraction of the surface Zn atoms lie close to a sulfur atom, because the two lattices are incommensurate (\cref{fig:S3}c,f).

\section{The interface dipole}
\label{sec:dipole}

\begin{figure}[H]
\centering
\includegraphics[width=\linewidth]{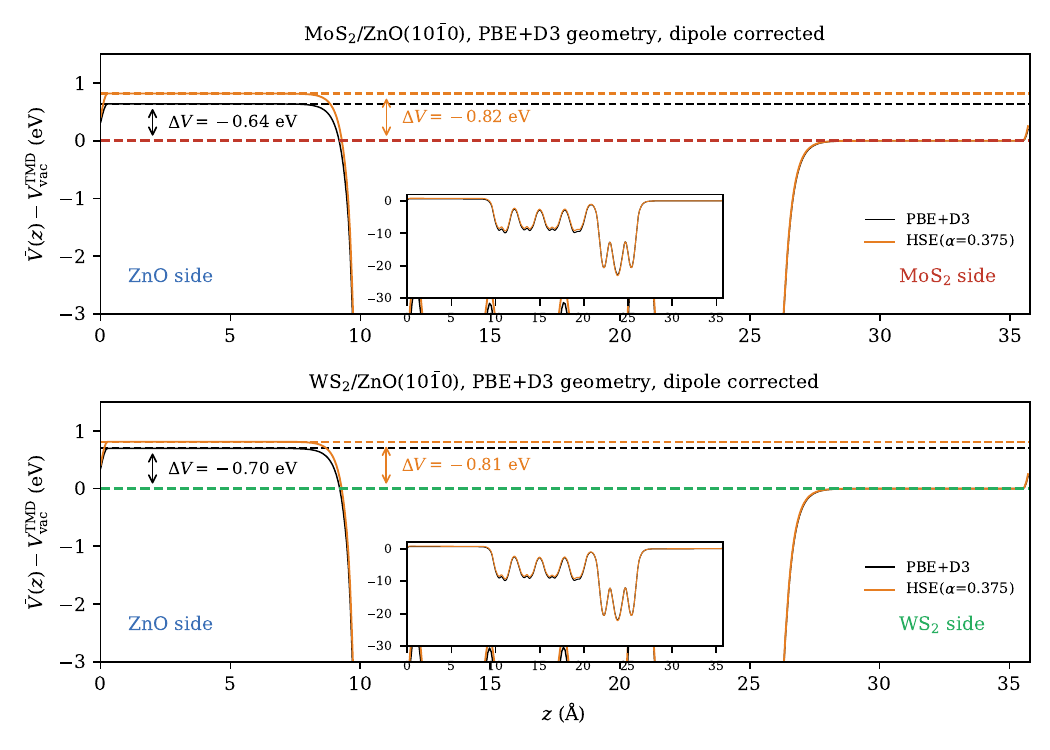}
\caption{\justifying Plane-averaged electrostatic potential of the dipole-corrected 330-atom interfaces at the PBE+D3 geometry (black: PBE+D3, orange: HSE(0.375)), relative to the vacuum level on the TMD side. The inset shows the full profile. The ZnO slab occupies $z \approx 10$-20\,\AA\ and the monolayer $z \approx 22$-25\,\AA. The vacuum level on the ZnO side lies $|\Delta V|$ above that on the TMD side.}
\label{fig:S4}
\end{figure}

\begin{table}[H]
\centering
\caption{\justifying Interface potential step $\Delta V$ (eV) at the PBE+D3 geometry.}
\label{tab:S5}
\vspace{0.3cm}
\begin{tabular}{lccc}
\toprule
interface & PBE+D3 & HSE06 & HSE(0.375)\\
\midrule
MoS$_2$/ZnO & $-0.64$ & $-0.76$ & $-0.82$\\
WS$_2$/ZnO & $-0.70$ & $-0.77$ & $-0.81$\\
\bottomrule
\end{tabular}
\end{table}

The negative sign means that the electrostatic energy of an electron on the TMD side is lowered with respect to the ZnO side, that is, an interface dipole with its positive end on the monolayer. This is the sign expected from the partial donation of sulfur lone-pair density into the empty dangling-bond orbitals of the threefold-coordinated surface Zn atoms, and from the reduction of the surface-dimer tilt, which reduces the outward dipole of the clean ZnO surface. With a magnitude of about 0.8\,eV the dipole is large on the scale of the band offsets and decides the type of the alignment. The hybrid functionals increase the magnitude of the dipole it by 0.1-0.2\,eV with respect to PBE+D3; a plausible but unverified reason is that in PBE the O-$2p$ surface state of ZnO is resonant with the valence band of the monolayer, so that the mixing of the two partly counteracts the S$\to$Zn donation, whereas the hybrid functional separates them in energy. HSE06 and HSE(0.375) agree to 0.06\,eV. We use HSE(0.375), the functional also used for ZnO, for all final offsets.

\section{Band offsets and the comparison of functionals}
\label{sec:offsets}

\begin{figure}[H]
\centering
\includegraphics[width=\linewidth]{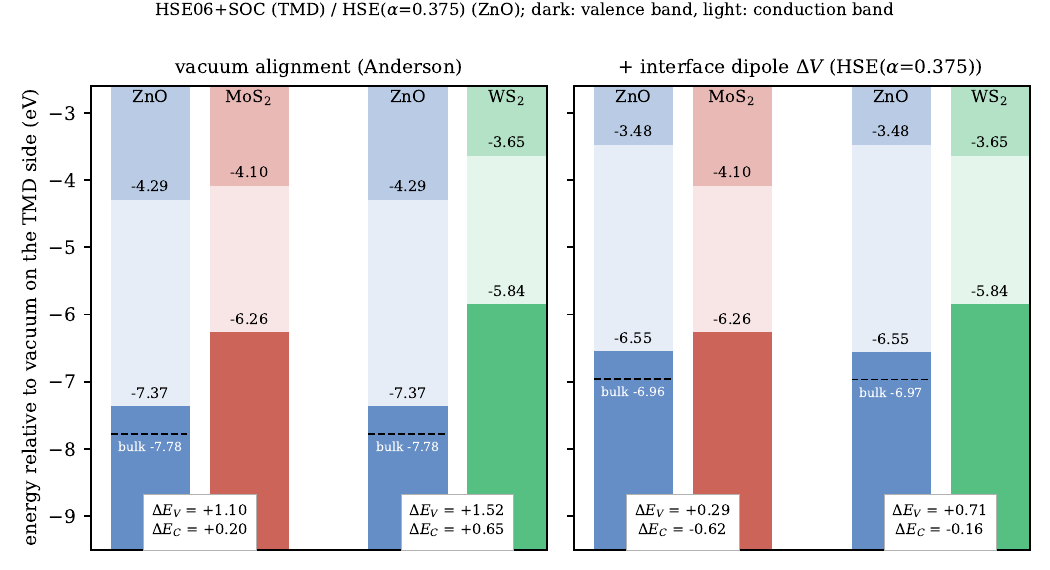}
\caption{\justifying Band alignment with HSE06+SOC for the monolayers and HSE(0.375) for ZnO. Left: vacuum-level (Anderson) alignment of the isolated systems. Right: including the interface dipole computed with HSE(0.375) at the PBE+D3 geometry. The ZnO energies are shifted by $-\Delta V$ and all energies refer to the vacuum level on the TMD side. Dark bars: valence bands; light bars: conduction bands; the dashed line in the ZnO valence band marks the bulk-like valence-band maximum below the O-$2p$ surface state.}
\label{fig:S5}
\end{figure}

\begin{table}[H]
\centering
\caption{\justifying Band offsets (eV) for all combinations of functionals, without ($\Delta V = 0$, Anderson rule) and with the interface dipole, which is always computed with the functional used for ZnO at the PBE+D3 geometry. Positive values place the TMD edge above the ZnO edge. Bold: recommended values.}
\label{tab:S6}
\vspace{0.3cm}
\begin{tabular}{lllcccc}
\toprule
 & & & \multicolumn{2}{c}{Anderson} & \multicolumn{2}{c}{+ dipole}\\
\cmidrule(lr){4-5}\cmidrule(lr){6-7}
 & TMD & ZnO & $\Delta E_V$ & $\Delta E_C$ & $\Delta E_V$ & $\Delta E_C$\\
\midrule
MoS$_2$ & PBE & PBE & $-0.08$ & $+1.01$ & $-0.72$ & $+0.37$\\
 & \textbf{HSE06+SOC} & \textbf{HSE(0.375)} & $+1.10$ & $+0.20$ & {\boldmath$+0.29$} & {\boldmath$-0.62$}\\
 & HSE(0.375)+SOC & HSE(0.375) & $+0.89$ & $+0.22$ & $+0.07$ & $-0.60$\\
 & HSE06+SOC & HSE06 & $+0.56$ & $+0.51$ & $-0.19$ & $-0.25$\\
\midrule
WS$_2$ & PBE & PBE & $+0.17$ & $+1.46$ & $-0.53$ & $+0.76$\\
 & \textbf{HSE06+SOC} & \textbf{HSE(0.375)} & $+1.52$ & $+0.65$ & {\boldmath$+0.71$} & {\boldmath$-0.16$}\\
 & HSE(0.375)+SOC & HSE(0.375) & $+1.32$ & $+0.70$ & $+0.50$ & $-0.12$\\
 & HSE06+SOC & HSE06 & $+0.98$ & $+0.96$ & $+0.21$ & $+0.19$\\
\bottomrule
\end{tabular}
\end{table}

With PBE the valence-band maximum is on ZnO for both interfaces ($\Delta E_V = -0.72$ and $-0.53$\,eV) and the conduction-band minimum on the monolayer, a consequence of the PBE ionization potential of ZnO and not of the interface. With the hybrid functionals the valence band of ZnO moves down by 1.5\,eV and lies below that of the monolayer, by 0.50 to 0.71\,eV for WS$_2$ and by 0.07 to 0.29\,eV for MoS$_2$, depending on the functional used for the monolayer. Without the dipole the alignment would be type II with electrons in ZnO; with the dipole both interfaces are type I. The valence-band offset of MoS$_2$/ZnO is the least certain number of this work: it is small with respect to the O-$2p$ surface state, 0.4\,eV larger with respect to the bulk-like valence band of ZnO, and its sign is reversed only if ZnO is described with HSE06 or PBE. Using HSE06 for ZnO, whose bulk-like gap of 2.60\,eV is 0.9\,eV too small, raises the ZnO valence band by 0.54\,eV and gives $\Delta E_V = -0.19$\,eV and $\Delta E_C = -0.25$\,eV for MoS$_2$/ZnO, that is, the valence-band maximum in ZnO and the conduction-band minimum in the monolayer, and $\Delta E_V = +0.21$\,eV, $\Delta E_C = +0.19$\,eV (type II with electrons in ZnO) for WS$_2$/ZnO. The ordering of the valence bands at this interface is therefore controlled by the description of the ZnO band edges, which is why we use the functional that reproduces the band gap, the Zn-$3d$ position and the measured ionization potential of the m-plane surface.

\begin{figure}[H]
\centering
\includegraphics[width=0.68\linewidth]{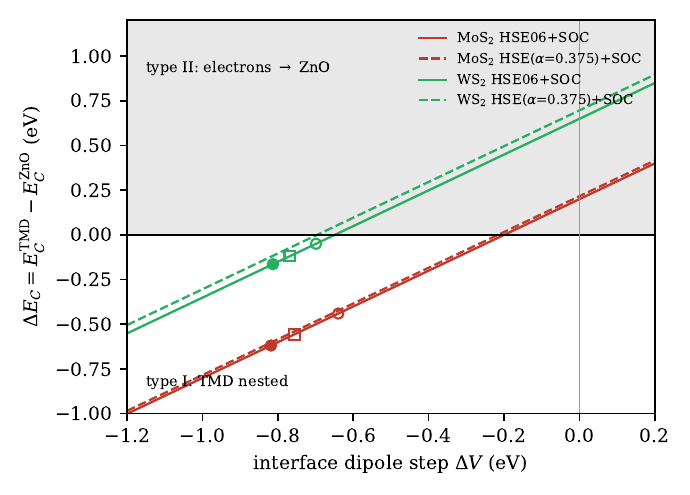}
\caption{\justifying Conduction-band offset as a function of the interface potential step $\Delta V$. Symbols: $\Delta V$ of the relaxed interfaces computed with HSE(0.375) (full circles), HSE06 (open squares) and PBE+D3 (open circles). In the gray region the alignment is type II with electrons transferred to ZnO. An increase of $\Delta V$ by about 0.16\,eV (WS$_2$) or 0.6\,eV (MoS$_2$) turns the interface into type II.}
\label{fig:S6}
\end{figure}

Since $\Delta E_C$ depends linearly on $\Delta V$, any modification of the interface that reduces the magnitude of the dipole by about 0.16\,eV (WS$_2$) or 0.6\,eV (MoS$_2$) produces a type-II alignment with electrons in ZnO. For comparison, the change of $\Delta V$ between a nearly and a fully relaxed geometry is 0.15\,eV for the WS$_2$ interface, the choice of functional changes it by up to 0.18\,eV, and hydroxylation of the same m-plane surface has been reported to shift the conduction-band offset by about 1.3\,eV.\cite{Sutarma.2026} Interfacial water, hydroxyl groups, oxygen vacancies and other adsorbates on the nanowire facet, as well as the band bending of an n-type nanowire, act in this direction. The type-II alignment with electrons in ZnO is therefore the alignment to be expected for the hydrated facet, while the ideal, dry facet is of type I.

The transfer of the holes, in contrast, is common to the two alignments and is insensitive to these details, because the valence-band offsets of 0.29\,eV (MoS$_2$) and 0.71\,eV (WS$_2$) exceed $k_\mathrm{B}T$ at room temperature by one to two orders of magnitude. What limits it is not the offset but the number of states available in the monolayer. With a hole effective mass of $0.6\,m_0$ in the two K valleys, considering only the upper spin branch, the valence density of states of MoS$_2$ is $2.5\times10^{14}$\,eV$^{-1}$cm$^{-2}$, so that filling the window of 0.29\,eV requires $7\times10^{13}$ holes per cm$^2$. For a nanowire of 250\,nm diameter this sheet density is supplied by a carrier density of $1.2\times10^{19}$\,cm$^{-3}$ if the whole perimeter is in contact and of $4.5\times10^{17}$\,cm$^{-3}$ for a contact 30\,nm wide, which brackets the density at which the extraction saturates. The positive charge accumulated in the monolayer, together with the electrons left behind in the nanowire, lowers the levels of the monolayer with respect to those of ZnO, so that the hole transfer reduces its own driving force and saturates. The same photovoltage shifts the conduction-band offset away from the type-II situation, so that illumination alone does not convert an ideal interface into type II.

The same counting applies to the conduction side, and for WS$_2$ it is the conduction side that limits the transfer. The windows are 0.29\,eV for holes and 0.62\,eV for electrons at the MoS$_2$ interface, and 0.71\,eV for holes and 0.16\,eV for electrons at the WS$_2$ interface, so that the bottleneck is the valence band in the first case and the conduction band in the second. The capacities are nevertheless of the same order, because the width of the window and the density of states compensate each other. Filling 0.16\,eV of the conduction band of WS$_2$ involves not only the two K valleys but also the six Q valleys, which the valley-resolved spin-orbit correction places within 3\,meV of K (\cref{sec:monolayers}) and which have a heavier mass, so that the conduction density of states is of the order of $10^{15}$\,eV$^{-1}$cm$^{-2}$ and the capacity of order $10^{14}$\,cm$^{-2}$. Filling 0.71\,eV of its valence band reaches the lower spin branch at K, 0.43\,eV below the maximum, and the $\Gamma$ valley, so that the capacity there also exceeds $10^{14}$\,cm$^{-2}$. All four numbers lie between $7\times10^{13}$ and a few $10^{14}$\,cm$^{-2}$, that is, between a few $10^{17}$ and about $10^{18}$\,cm$^{-3}$ in the nanowire for the contact geometry of the experiment, below the carrier density at threshold in every case.

Electrostatics couples the two flows. If the holes run ahead, as they do at the WS$_2$ interface where their offset is four times the conduction offset, the positive charge left in the monolayer lowers its levels with respect to those of ZnO, which reduces the valence offset and deepens the conduction offset, so that the imbalance suppresses itself and the capture proceeds close to charge neutral, pair by pair. The space-charge argument of the preceding paragraph therefore describes the transient, and in the balanced steady state the transfer is limited by Pauli blocking of the shallower of the two windows.

The type-I alignment of the ideal interface therefore already accounts for the two experimental observations, the quenching of the emission and the increase of the threshold, and neither of them requires a type-II alignment. The blue shift of the lasing modes is not an unambiguous fingerprint of band filling either, since band filling displaces the whole gain spectrum, including its high-energy edge, whereas the measured mode envelope keeps its high-energy cutoff and loses modes on its low-energy flank, which is the signature of a wavelength-dependent loss.

\section{Validation: direct hybrid calculations of the 330-atom heterostructures}
\label{sec:direct}

As a check of the three-step scheme, both interfaces were also computed self-consistently with HSE(0.375) and with HSE06 ($\Gamma$ point, dipole correction, PBE+D3 geometry, no spin-orbit coupling). These are the calculations that provide the hybrid $\Delta V$ of \cref{sec:dipole}. Their band edges can be compared with the three-step result evaluated with the same functional and the same $\Delta V$, which removes the dipole from the comparison.

Two points require attention. First, the K point of the monolayer does not fold onto $\Gamma$ of the supercell: it lies at $(0,\tfrac{1}{3})$ of the rectangular cell and folds onto $(0,\tfrac{2}{3})$ of the $3\times5$ supercell, because 5 periods along $y$ are not a multiple of three. A $\Gamma$-only calculation of the heterostructure therefore misses the TMD band edges. The $\Gamma$ point of the supercell samples exactly the $3\times5$ $\Gamma$-centered mesh of the rectangular cell, including the sum over $q$ in the Fock exchange, so the correction can be obtained from the isolated strained monolayer computed on that mesh and on a $6\times15$ mesh that contains K. It moves the valence-band maximum up by 0.11-0.18\,eV and the conduction-band minimum down by 0.09-0.16\,eV. A non-self-consistent PBE+D3 calculation of the interfaces at the supercell point $(0,\tfrac{1}{3},0)$ confirms that the same correction applies in the heterostructure. Second, when the valence band of the monolayer at its own $\Gamma$ point lies within 0.10-0.15\,eV of the O-$2p$ surface state of ZnO, the two mix and the TMD-majority combination is pushed up by up to 0.2\,eV. Adding the K-valley correction to such a state would count the shift twice. The valence-band maximum of the monolayer is therefore taken as the higher of the hybridized $\Gamma$-derived state and of the corrected K-valley maximum.

\begin{table}[H]
\centering
\caption{\justifying Direct band offsets (eV) of the 330-atom heterostructures compared with the three-step result for the same functional and the same $\Delta V$ (strained isolated monolayer on the K-containing mesh, 8-layer ZnO slab, no spin-orbit coupling in either). $\delta K_{v,c}$: K-valley correction of the TMD band edges. $\Delta\mathrm{EA}^\mathrm{ZnO}$ and $\Delta\mathrm{EA}^\mathrm{TMD}$: electron affinity in the heterostructure minus that of the isolated system. Last column: valley of the TMD valence-band maximum in the heterostructure ($\Gamma$ denotes the hybridized $\Gamma$-derived state).}
\label{tab:S7}
\small
\vspace{0.3cm}
\begin{tabular}{llccccccccc}
\toprule
 & functional & $\delta K_v$ & $\delta K_c$ & \multicolumn{2}{c}{$\Delta E_V$} & \multicolumn{2}{c}{$\Delta E_C$} & $\Delta\mathrm{EA}^\mathrm{ZnO}$ & $\Delta\mathrm{EA}^\mathrm{TMD}$ & VBM\\
\cmidrule(lr){5-6}\cmidrule(lr){7-8}
 & & & & direct & 3-step & direct & 3-step & & & \\
\midrule
MoS$_2$ & PBE+D3 & $+0.11$ & $-0.09$ & $-0.68$ & $-0.67$ & $+0.17$ & $+0.52$ & $-0.31$ & $+0.03$ & K\\
 & HSE(0.375) & $+0.15$ & $-0.16$ & $+0.11$ & $+0.05$ & $-0.66$ & $-0.36$ & $-0.26$ & $+0.04$ & $\Gamma$\\
 & HSE06 & $+0.14$ & $-0.14$ & $-0.20$ & $-0.21$ & $-0.33$ & $-0.02$ & $-0.27$ & $+0.03$ & K\\
\midrule
WS$_2$ & PBE+D3 & $+0.15$ & $-0.12$ & $-0.51$ & $-0.49$ & $+0.39$ & $+0.75$ & $-0.30$ & $+0.06$ & K\\
 & HSE(0.375) & $+0.18$ & $-0.12$ & $+0.35$ & $+0.33$ & $-0.21$ & $+0.07$ & $-0.24$ & $+0.03$ & K\\
 & HSE06 & $+0.17$ & $-0.12$ & $+0.06$ & $+0.04$ & $+0.02$ & $+0.32$ & $-0.26$ & $+0.04$ & $\Gamma$\\
\bottomrule
\end{tabular}
\end{table}

The valence-band offsets agree to 0.06\,eV or better for all six combinations, so the direct calculations confirm the three-step valence-band offsets, including the small MoS$_2$ value. The direct conduction-band offsets are more negative than the three-step ones by 0.28-0.36\,eV in every case, almost entirely because of the ZnO conduction band: in the heterostructure its electron affinity is 0.24-0.31\,eV smaller than in the 8-layer slab, since the delocalized Zn-$4s$ conduction-band minimum is confined in the four ZnO layers of the interface cell, whereas the conduction band of the monolayer is within 0.06\,eV of that of the isolated strained layer. A nanowire facet is the surface of a thick crystal, so the 8-layer slab used in the three-step scheme is the appropriate reference and the direct calculations overestimate the conduction-band offset in favor of type I. The direct calculations give no indication of interface states or of charge transfer beyond what is already contained in $\Delta V$.

\begin{figure}[H]
\centering
\includegraphics[width=\linewidth]{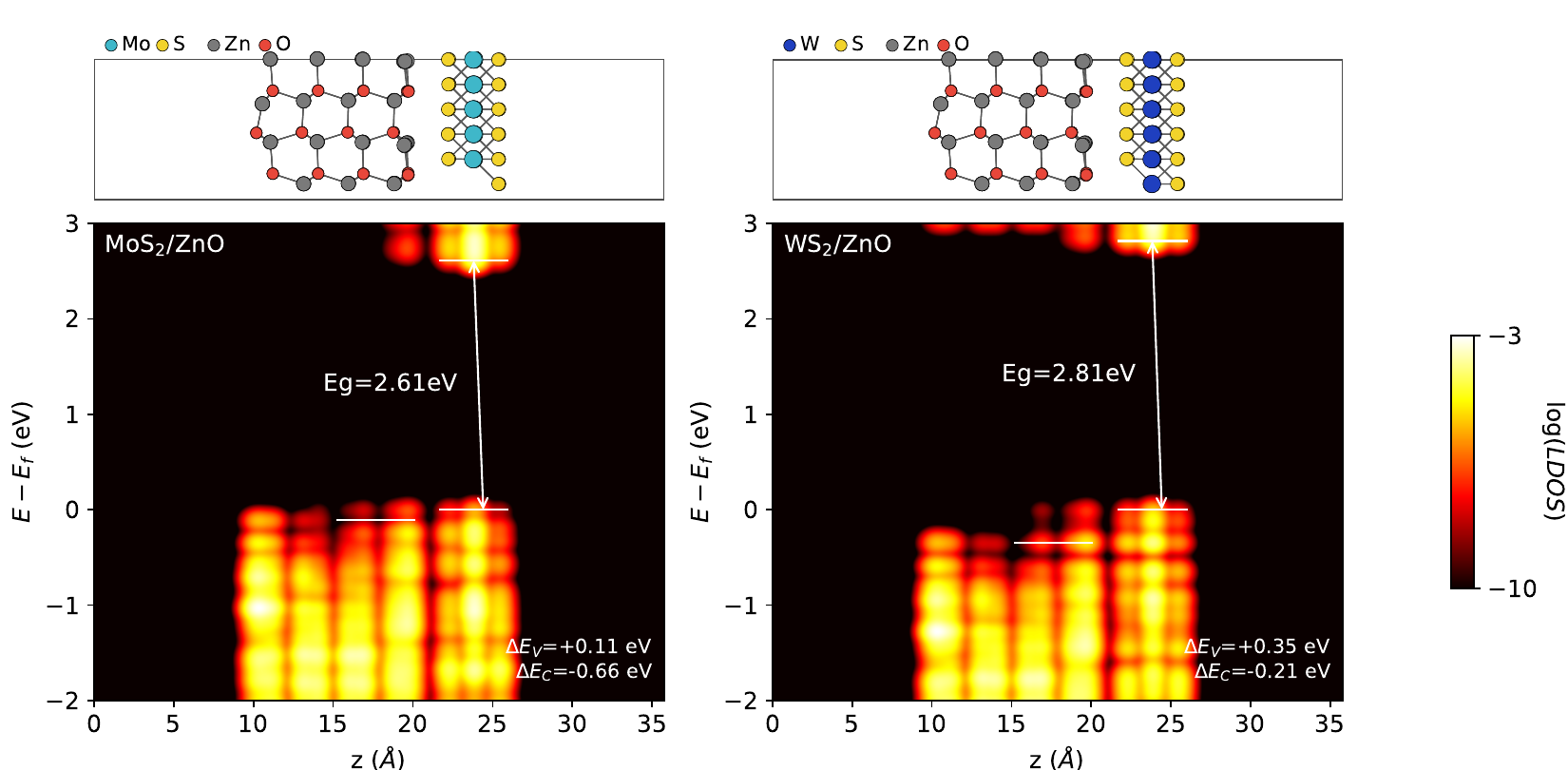}\\[1ex]
\includegraphics[width=\linewidth]{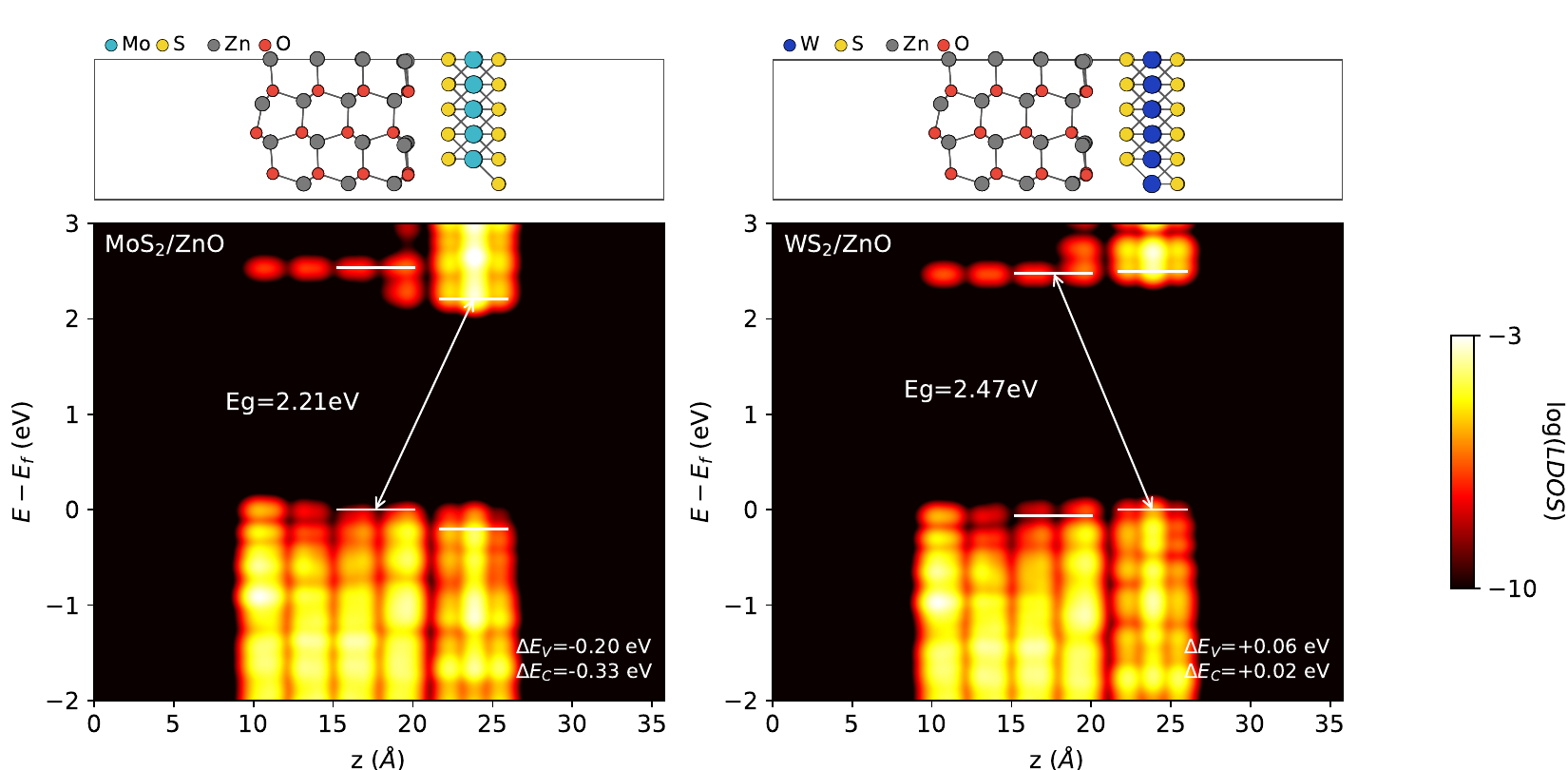}
\caption{\justifying Layer-resolved log(LDOS) maps of the direct HSE(0.375) (top) and HSE06 (bottom) calculations of the heterostructures, in the style of Fig.~3e,f of the main text. The TMD part of the TMD-majority bands is moved to the K valley as described in the text; no spin-orbit coupling. For HSE(0.375) the conduction band of the four-layer ZnO slab lies above the plotted window. Because of the thin slab the conduction-band offsets of these maps are about 0.3\,eV too negative.}
\label{fig:S7}
\end{figure}

\section{Layer-resolved densities of states}
\label{sec:ldos}

The maps of Fig.~3e,f and of \cref{fig:S7,fig:S9} are built from the \texttt{PROCAR} ion projections $w_{na}$ of the heterostructure as
\begin{equation}
\mathrm{LDOS}(z,E) = \frac{2}{A}\sum_n\sum_a \tilde{w}_{na}\, g_{\sigma_E}(E-\varepsilon_{na})\, g_{\sigma_z}(z-z_a),
\label{eq:ldos}
\end{equation}
with $\tilde{w}_{na}$ the projections normalized to one per band, $A$ the cell area and Gaussians of width $\sigma_E = 0.06$\,eV and $\sigma_z = 0.45$\,\AA. In the maps of Fig.~3e,f of the main text the part of band $n$ on subsystem $X$ (the monolayer or ZnO) is shifted by a constant chosen so that the band edges of $X$ move from their PBE+D3 positions in the heterostructure to the final values of \cref{tab:S6}. A band edge of a subsystem is the extremal band carrying at least 30\% of its weight in that subsystem. A band with less than 30\% of its weight on one side is shifted rigidly with the other side, so that the evanescent tail of a state stays at the energy of the state. This applies only to the ZnO atoms within 5\,\AA\ of the interface, since a tail decays within a few ångströms, whereas weight further inside the slab always follows the ZnO alignment. The rigid shift does not remove the hybridization that is present in the PBE wave functions: in WS$_2$/ZnO the $\Gamma$ valence state of the monolayer, resonant with the ZnO valence band in PBE, keeps a weak tail on the interfacial ZnO layer after it is moved into the gap of ZnO.

\begin{figure}[H]
\centering
\includegraphics[width=\linewidth]{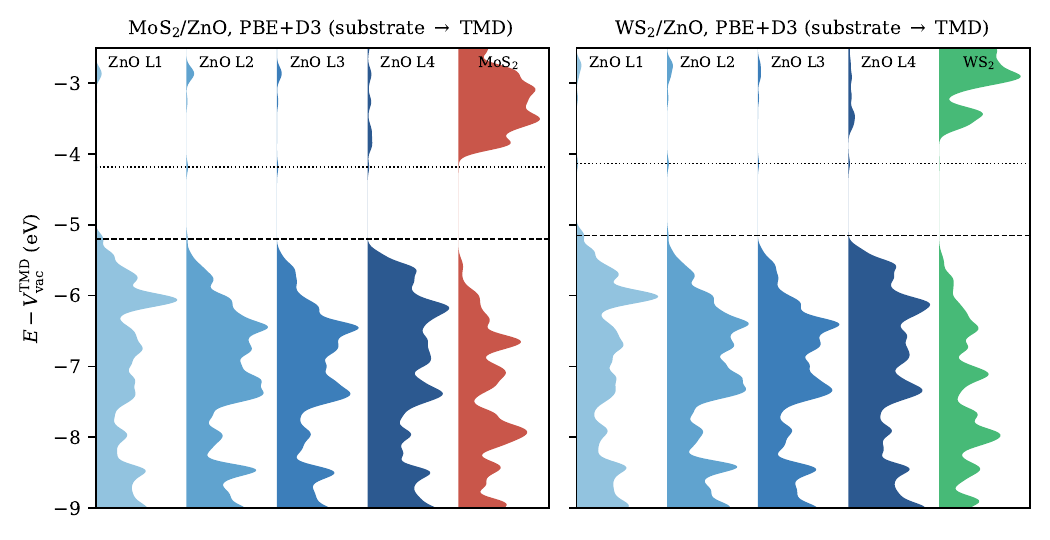}
\caption{\justifying Layer-projected density of states of the PBE+D3 heterostructures ($\Gamma$ point, Gaussian broadening 0.08\,eV, each panel normalized to its maximum), from the bottom ZnO layer (L1, facing the vacuum) through the interfacial ZnO layer (L4) to the monolayer. Energies refer to the vacuum level on the TMD side; dashed and dotted lines are the valence-band maximum and the conduction-band minimum of the heterostructure. The small weight of the Zn-$4s$ conduction band is an artifact of the PAW projection spheres, which capture little of this delocalized state.}
\label{fig:S8}
\end{figure}

\begin{figure}[H]
\centering
\includegraphics[width=\linewidth]{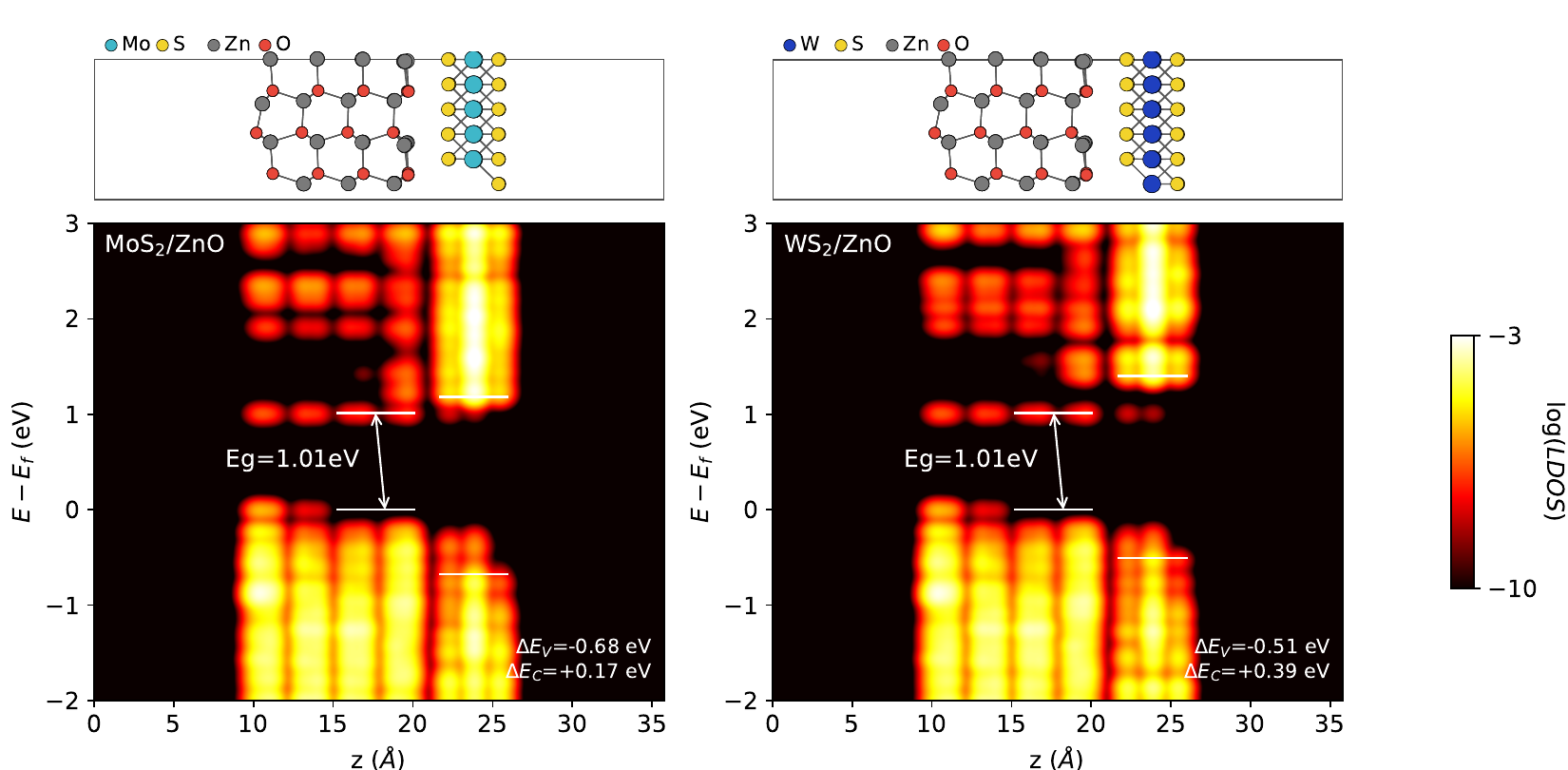}
\caption{\justifying Layer-resolved log(LDOS) maps of the PBE+D3 heterostructures, with the TMD band edges moved from $\Gamma$ of the supercell to the K valley. With a semilocal functional the valence-band maximum is on ZnO, on the surface states of both ZnO surfaces, and the conduction-band minimum on the monolayer. Comparison with Fig.~3e,f of the main text shows that this ordering is a consequence of the PBE band edges of ZnO and not of the interface.}
\label{fig:S9}
\end{figure}

In the PBE heterostructures the valence-band maximum lies 5.20\,eV (MoS$_2$) and 5.15\,eV (WS$_2$) below the vacuum level on the TMD side, while the three-step expression evaluated with the isolated four-layer slab predicts 5.18 and 5.12\,eV: the alignment scheme reproduces the direct heterostructure calculation to 0.03\,eV.

\section{Error budget}
\label{sec:errors}

\begin{table}[H]
\centering
\caption{\justifying Estimated uncertainties of the band offsets.}
\label{tab:S8}
\small
\vspace{0.3cm}
\begin{tabular}{>{\raggedright}p{0.24\linewidth} >{\raggedright}p{0.52\linewidth} >{\raggedright\arraybackslash}p{0.16\linewidth}}
\toprule
source & test & effect (eV)\\
\midrule
direct versus three-step & valence-band maximum of the PBE heterostructure, and $\Delta E_V$ of the direct PBE, HSE06 and HSE(0.375) heterostructures, against the three-step expression & 0.06\\
ZnO slab thickness for $\Delta E_C$ & electron affinity of ZnO in the four-layer heterostructure against the 8-layer slab (affects the direct cells only) & 0.24-0.31\\
slab thickness for the IP & 4 against 8 layers (PBE) & 0.005\\
$k$ sampling of $\Delta V$ & $\Gamma$ against $2\times2\times1$ at 400\,eV & 0.032\\
cutoff of $\Delta V$ & 400\,eV/Normal against 500\,eV/Accurate & 0.003\\
geometry & PBE+D3 $\Delta V$ at $F_\mathrm{max}\approx0.1$ against $<0.02$\,eV/\AA & $\le 0.15$\\
vacuum level & flatness of the potential in the averaging window & $<0.001$\\
functional for the monolayer & HSE06+SOC against HSE(0.375)+SOC & 0.2 ($\Delta E_V$), 0.05 ($\Delta E_C$)\\
functional for $\Delta V$ & HSE(0.375) against HSE06 & 0.06\\
strain of the monolayer in the supercell & about 1\%, enters only through $\Delta V$ & $<0.05$\\
spin-orbit coupling & PBE correction instead of a hybrid+SOC calculation & $<0.05$\\
\bottomrule
\end{tabular}
\end{table}

The dominant uncertainty is the interface dipole, through its sensitivity to the final geometry (0.15\,eV for WS$_2$ between a nearly and a fully relaxed structure) and to the functional (0.18\,eV between PBE+D3 and HSE(0.375)), both comparable with the conduction-band offset of WS$_2$/ZnO. The valence-band offset of WS$_2$/ZnO is positive by at least 0.5\,eV in every combination of hybrid functionals. The offset of MoS$_2$/ZnO is positive but small with respect to the ZnO surface state and 0.4\,eV larger with respect to the bulk-like valence band. Many-body corrections beyond hybrid DFT and the large exciton binding energies of the monolayers (0.2-0.9\,eV, depending on the dielectric environment) affect the optical, not the electronic, alignment discussed here.

\section{Comparison with previous calculations and with photoemission}
\label{sec:comparison}

\begin{table}[H]
\centering
\caption{\justifying Comparison for WS$_2$ on m-plane ZnO. The values of Ref.~\cite{Sutarma.2026} for the offsets are read from the band diagram of that work, which aligns the vacuum levels of its models. The sum of the two offsets reproduces the difference of the quoted gaps to 0.01\,eV. The last column is our direct HSE(0.375) calculation of the 330-atom heterostructure (\cref{tab:S7}) with the spin-orbit corrections of \cref{tab:S2} added, which is the calculation directly comparable with Ref.~\cite{Sutarma.2026}.}
\label{tab:S9}
\small
\begin{tabular}{>{\raggedright}p{0.2\linewidth} >{\raggedright}p{0.22\linewidth} >{\raggedright}p{0.24\linewidth} >{\raggedright\arraybackslash}p{0.24\linewidth}}
\toprule
 & this work (recommended) & Ref.~\cite{Sutarma.2026}, pristine & this work, same setup as Ref.~\cite{Sutarma.2026}\\
\midrule
functional for the TMD & HSE06+SOC & HSE(0.375)+SOC & HSE(0.375), SOC added a posteriori\\
ZnO reference & 8-layer symmetric slab & passivated slab, about 6 layers & 4 layers in the interface cell\\
$d_\mathrm{Zn-S}$ (\AA) & 2.66 & 2.615 & 2.66\\
$E_g$(ZnO) (eV) & 3.49 & 3.42 & --\\
$E_g$(WS$_2$) (eV) & 2.20 & 2.52 isolated, 2.41 at the interface & 2.45\\
$\Delta E_V$ (eV) & $+0.71$ & $+0.57$ & $+0.56$\\
$\Delta E_C$ (eV) & $-0.16$ & $-0.45$ & $-0.28$\\
alignment & type I & type I & type I\\
\bottomrule
\end{tabular}
\end{table}

The only previous calculation of this interface, Ref.~\cite{Sutarma.2026}, treats pristine WS$_2$/\ZnOm\ with the same hybrid functional that we use for ZnO and reads the band edges from element-projected band structures of a single supercell, without separating the interface dipole. The two approaches agree on the alignment type and, once the same functional and the same route are used, on the valence-band offset to 0.01\,eV ($+0.56$ against $+0.57$\,eV). The conduction-band offsets differ by 0.17\,eV, which is within the uncertainty of the dipole in \cref{tab:S8}. Our recommended values differ from those of Ref.~\cite{Sutarma.2026} for two reasons that act in known directions: HSE06 instead of $\alpha = 0.375$ for the monolayer raises $\Delta E_V$ by about 0.2\,eV, because a single fraction of exact exchange cannot describe both a wide-gap oxide and a two-dimensional semiconductor with strongly non-local screening, and the eight-layer ZnO reference removes the quantum confinement of the Zn-$4s$ conduction band that makes $\Delta E_C$ too negative in a thin slab. Ref.~\cite{Sutarma.2026} also finds that hydroxylation of the same surface converts the junction into type II with electrons in ZnO, shifting the conduction-band offset by about 1.3\,eV, which is the mechanism discussed in \cref{sec:offsets}. No calculation of MoS$_2$ on the m-plane facet is available for comparison.

Published calculations of TMD monolayers on graphene-like ZnO monolayers \cite{Chen.2026} describe a different system: those slabs have no bulk ZnO band edges and a different electronic structure, and above three layers they become polar $(0001)$ slabs. They are therefore not comparable with a nanowire side facet, although they also show that the alignment is sensitive to strain. The XPS/UPS study of Ref.~\cite{Eobaldt.2022} was performed on MoS$_2$ transferred onto c-plane ZnO$(0001)$ single crystals measured without further cleaning, and deduces a type-II alignment with electrons in ZnO. The c-plane is polar and its ionization potential differs by 0.9\,eV from that of the m-plane facet (\cref{sec:zno}), so those values cannot be transferred to the nanowire side facets. The comparison is meaningful only at the level of the transfer direction, which our calculations reproduce once the magnitude of the interface dipole is reduced, for instance by the hydroxylation expected for air-exposed surfaces. The samples of Ref.~\cite{Eobaldt.2022} were measured without any cleaning procedure, in order to reproduce the surface conditions of the lasing experiments, and therefore correspond to the hydrated rather than to the ideal facet. The type-II alignment that they report is consistent with the hydroxylated model of Ref.~\cite{Sutarma.2026} and not in contradiction with the type-I alignment that we obtain for the pristine interface.

\printbibliography